\documentclass[preprint,12pt]{elsarticle}

\usepackage{psfrag}
\usepackage{amsmath,amssymb,amsthm}
\usepackage{color}
\usepackage{amsfonts}
\usepackage{graphicx}
\usepackage{bm}
\usepackage{labelfig}
\newcommand{\eps}{\varepsilon}
\newcommand{\beps}{\bm{\varepsilon}}
\newcommand{\bepsscript}{\bm{\varepsilon}}

\newcommand{\bsig}{\mbox{\boldmath $\sigma$}}

\newcommand{\btau}{\mbox{\boldmath $\tau$}}

\newcommand{\tr}{\mathrm{tr}}
\newcommand{\blambda}{\mbox{\boldmath $\lambda$}}

\newcommand{\bA}{\mathbf{A}}
\newcommand{\bB}{\mathbf{B}}
\newcommand{\bC}{\mathbf{C}}

\newcommand{\bE}{\mathbf{E}}
\newcommand{\bG}{\mathbf{G}}

\newcommand{\bI}{\mathbf{I}}
\newcommand{\bK}{\mathbf{K}}
\newcommand{\bL}{\mathbf{L}}

\newcommand{\bT}{\mathbf{T}}

\newcommand{\ba}{\mathbf{a}}

\newcommand{\be}{\mathbf{e}}

\newcommand{\bk}{\mathbf{k}}

\newcommand{\bn}{\mathbf{n}}
\newcommand{\bq}{\mathbf{q}}
\newcommand{\bu}{\mathbf{u}}
\newcommand{\bt}{\mathbf{t}}
\newcommand{\bx}{\mathbf{x}}

\newcommand{\bM}{\mathbf{M}}

\newcommand{\obeps}{\overline{\beps}}

\def\sk#1{[\![#1]\!]}

\newcommand{\Cof}{\textrm{Cof\,}}

\DeclareMathOperator*{\noSum}{\mathchoice%
{\mbox{\footnotesize$\displaystyle\sum$}\mkern-16.5mu\raise-0.75ex\hbox{\mbox{\LARGE/}}\mkern2mu}%
{\mbox{\small$\textstyle\sum$}\mkern-13.5mu\raise-0.5ex\hbox{\mbox{\Large/}}\mkern3mu}%
{\mbox{\small$\scriptstyle\sum$}\mkern-12mu\raise-0.33ex\hbox{\mbox{/}}\mkern2mu}%
{\mbox{\small$\scriptscriptstyle\sum$}\mkern-11mu\raise-0.26ex\hbox{\mbox{\footnotesize/}}\mkern2mu}
} 
\newtheorem{proposition}{Proposition}

\journal{International Journal of Engineering Science}

\begin{document}

\begin{frontmatter}

\synctex=1

\title{
Phase transformations surfaces and exact energy lower bounds
}

\author[SPbSTU]{Mikhail~A.~Antimonov}
\ead{mike.antimonov@gmail.com}
\author[UUtah]{Andrej~Cherkaev}
\ead{cherk@math.utah.edu}
\author[SPbSTU,IPME,SPbSU]{Alexander~B.~Freidin\corref{SPbSTU,IPME,SPbSU}}
\cortext[SPbSTU,IPME,SPbSU]{Corresponding author}
\ead{alexander.freidin@gmail.com}

\address[SPbSTU]{Peter the Great
St. Petersburg Polytechnic University\\
Polytechnicheskaya Street, 29, Saint Petersburg, 195251, Russia}
\address[UUtah]{University of Utah\\
155 S 1400 E, Salt Lake City, UT, 84112, USA}
\address[IPME]{Institute for Problems in Mechanical Engineering of Russian Academy of Sciences\\
Bolshoy Pr., 61, V.O., Saint Petersburg, 199178, Russia}
\address[SPbSU]{St. Petersburg State University\\
Universitetsky Pr., 28, Peterhof, Saint Petersburg, 198504, Russia\\
\vspace{10mm}
Dedicated to Professor Sergey Kanaun on the occasion of his 70th birthday.}

\begin{abstract}

The paper investigates two-phase microstructures of optimal 3D composites that store minimal elastic energy in a given strain field. The composite is made of two linear isotropic materials which differ in elastic moduli and self-strains. We find optimal microstructures for all values of external strains and volume fractions of components. This study continues research by Gibiansky and Cherkaev \cite{gibiansky1987microstructures,GibianskyCherkaev1997} and Chenchiah and Bhattacharya \cite{Bhattacharya2008}. In the present paper we demonstrate that the energy is minimized by that laminates of various ranks. Optimal structures are either simple laminates that are codirected with external eigenstrain directions, or inclined laminates, direct and skew second-rank laminates and third-rank laminates. These results are applied for description of direct and reverse transformations limit surfaces in a strain space for elastic solids undergoing phase transformations of martensite type. The surfaces are computed as the values of external strains at which the optimal volume fraction of one of the phases tends to zero. Finally, we compare the transformation surfaces with the envelopes of the nucleation surfaces constructed earlier for nuclei of various geometries (planar layers, elliptical cylinders, ellipsoids). We show the energy equivalence of the cylinders and direct second-rank-laminates, ellipsoids and third-rank laminates. We note that skew second-rank laminates make the nucleation surface convex function of external strain, and they do not correspond to any of the mentioned nuclei.

\end{abstract}


\end{frontmatter}

\section{Introduction}

In the present paper, we relate two problems which initially arise from different branches of mechanics of materials: construction of transformation surfaces for strain-induced phase transitions and optimal design of 3D-composites in the sense of minimizing its energy. Both problems are formulated as a variational problem about a structure of two-phase elastic composite of minimal energy, they are solved by the same formalism. The previous paper \cite{gibiansky1987microstructures}, translated as \cite{GibianskyCherkaev1997}, considered the optimal composite   in an asymptotic case when the compliance of one phase was  zero; and the paper \cite{Bhattacharya2008} deals with a similar variational problem and contains results for two-phase composites in two and three dimensions, however, some optimal structures were not specified there.

Phase transformations of martensite type are characterized by a priori unknown interfaces which divide a body into domains occupied by different phases; these are accompanied by transformation strains, jumps of elastic moduli and a jump of a chemical energy  (see, e.g., \cite{Boiko1994, Lexcellent2013}, and reference therein).
The transformation cannot occur until the external strain attains the transformation limit surface.

One of the  approaches to the transformation surface construction can be based on a semi-inverse method.  According to this approach the new phase nucleus shape is prescribed (layers, ellipsoids, elliptical cylinders), and external strains at which the boundary of such a nucleus can satisfy the local thermodynamic equilibrium condition (the Maxwell relation) are  found. Then the geometric parameters of the nucleus  can be found in dependence on the external strains  (see, e.g., \cite{Grinfeld1991,Kublanov1988,Freidin1989,KaganovaRoitburd1988, FreidinSharipova2006,Freidin2007,Antimonov2009,AntimonovNTV-2010Cylinder, Freidin2009}). Then some kind of a transformation surface can be constructed as an envelope of the surfaces which correspond to external strains at which the appearance of different types of new phase nuclei becomes possible.

Such an approach  allows  to specify the shape and orientation of new phase nucleus
in dependence on strain state and to study local strains. However, the local equilibrium conditions are only  necessary conditions for
the energy  minimum and an additional stability analysis is needed. It was
shown that the instability of such two-phase deformations with respect to  various smooth perturbations of the interface was not found if strains at the interfaces  corresponded to the external boundaries of so-called phase transitions zones (see, e.g., \cite{FuFreidin2004,EremeyevFreidinSharipovaPMM2007}). The phase transition zone (PTZ) is formed in a strain space by all strains which can satisfy the local equilibrium conditions and allows to describe locally all equilibrium interfaces feasible in a given material. The concept of the PTZ  was offered in \cite{FreidinChiskis1}  and its construction was considered  for finite and small strains \cite{FreidinChiskis2,MorozovFreidin1998,FreidinUniv1999,FreidinVilchSharTAM2002, FreidinFuSharVil2006,FreidinSharipova2006}.  The external PTZ-boundaries are the surfaces of the nucleation of new phase plane layers. In the papers \cite{GrabovskyTruskinovsky2011,GrabovskyTruskinovsky2013} it was proved  that belonging strains to the external PTZ-boundaries is a necessary stability condition.

But even if these necessary conditions are satisfied by the choice of the geometry of a new phase domain and instability of the solution chosen is not found,  one cannot be sure that the solution corresponds to the energy minimizer.
We also note that the variational problem in the case of phase transformation has a two-wells nonconvex Lagrangian, therefore, the conventional variational technique is not applicable; the solution is characterized by a microstructure of mixed phases with a generalized boundary between them. Such solutions are investigated by constructing the minimizing sequences and establishing a lower bound for the energy \cite{BallJames1987,ball1989fine}.

 That is why, accepting that the transformation cannot start if two-phase microstructure appear\-ance do not lead to energy relaxation in comparison with one-phase states, we develop the   approach based on the construction of the exact energy lower bounds of two-phase composites. The energy functional for the case of a given external strain  and temperature is the Helmholtz free energy. We assume that the phases are linear-elastic and, for simplicity, isotropic.

The transformation surface construction includes several  steps.
We find the structure of two-phase composites that stores minimal energy at a given external strain, using different constructions for upper and lower bounds of the energy and observe that these bounds coincide. This approach was used in most papers on the subject starting with \cite{lurie1983optimal,BallJames1987,lurie1988certain}. In most papers on optimal  3D elastic composites, the minimum of the
complementary energy (dual or stress energy) was considered, since it leads to the strongest structures, see e.g. \cite{francfort1995fourth,cherkaev1996optimal,diaz1997optimal,cherkaev1998stable, allaire1997shape}. We, however, consider the minimum of the strain energy (the weakest composites), as it is required by the Gibbs principle.

Both upper and lower bounds require a nontrivial search for optimal parameters of the estimates. We use them simultaneously, finding the hints for values of the parameters of the lower estimate from the upper estimate and vice versa. A similar strategy was exploited  starting with \cite{gibiansky1987microstructures,Milton1986} (see also \cite{GibianskyCherkaev1997} and monographs \cite{Cherkaev2000, Milton2002}) and has been actively elaborated for the estimations of electric, thermal and elastic  properties of composite materials basing on  the translation method \cite{Lurie1982, Milton1990,Firoozye1991,Seregin1996,Grabovsky09041996,GibianskyCherkaev1997, Avellaneda19961179}.

We find a two-phase composite of minimal strain energy at given volume fractions of the components.   We start with considering the specific microstructures namely  laminates of different ranks and we minimize the energy in this class of structures obtaining the upper bound for the minimizing energy.
Note that obtaining exact energy estimations with the use of layered microstructures became standard procedure in the analysis of optimal microstructures \cite{GibianskyCherkaev1997, Bhattacharya2008, Allaire-1999, Avellaneda19961179, Milton1986, Kohn1991,cherkaev1996extremal,albin2007multiphase,cherkaev2011optimal}.

Then
we construct a lower translation bound for the composite energy (see, e.g., the monographs \cite{Cherkaev2000,Milton2002} and reference therein). The translation bound does not depend on microstructure and it  may or may not be  attained.
However,  the  direct energy calculation  shows that the energy of the optimal laminates coincides with the polyconvex energy envelope.  The  lower bound depends on so-called translation parameters that should be optimally adjusted. The translation method  provides optimal strains in the materials, which are utilized when we find optimal microstructures among laminates. This way, translation bound and optimal laminates' energy are coupled and allow for the definition of optimal translation parameters and optimal parameters of laminates. Here we follow the approach developed in \cite{cherkaev2011optimal,cherkaev2013three} for optimal structures of multimaterial mixtures.

Then we further minimize the lower bound with respect to the volume fractions and find all strains at which the minimizer corresponds to the limiting values (zero or one) of the volume fraction of one of the phases. The strains that correspond to  these values form the direct and the reverse transformation limit surfaces, respectively. Two-phase structures
cannot have lower energy than the energy of one-phase state until the limit surface is reached.  Such domains of pure materials and optimal mixtures in the context of structural optimization were described in \cite{cherkaev2004detecting1} and \cite{cherkaev2004detecting2} for 2D and 3D strongest elastic composites, respectively, and in \cite{briggs2015note} for 2D three-material  strongest elastic composites.

Finally, we compare the limit
surfaces constructed with  nucleation surfaces for ellipsoidal, cylindrical and planar nuclei. The coincidence of the surfaces indicates the fact that layered microstructures are energy  equivalent to microstructures with smooth interfaces (layers, elliptical cylinders, ellipsoids).

\section{Problem statement}

In this section, we formulate the problem of minimal Helmholtz free energy for a phase transition between two isotropic elastic phases and introduce notations. The formulation generally follows the variational approach presented in the earlier papers \cite{lurie1988certain,ball1989fine,Grinfeld1991}).

\paragraph{Geometry and elasticity} Consider a periodic structure in three-dimensional space $ \mathbb{R}^3$. A periodic cell is a unit cube  $${\Omega=\{x:0\leq x_i\leq1,i=\overline{1,3}\}},$$ $\text{meas}(\Omega)=1$. Assume that $\Omega$ is divided into two parts  $\Omega_-$ and $\Omega_+$ with volumes $\text{meas}(\Omega_-)=m_-$ and $\text{meas}(\Omega_+)=m_+$; obviously, $m_- + m_+=1$. We denote the interface between subregions as $\Gamma$ and introduce  the characteristic function $H(\bx)$ of the subregion $\Omega_+$
\begin{gather*}
H(\bx)=\left\{
	\begin{array}{l}
	1, \quad \bx\in \Omega_+,\\
	0, \quad \bx\in \Omega_-.
	\end{array}
\right.
\end{gather*}

 Assume that these parts are  occupied by  linear elastic phases with the elasticity tensors $\bC_{-}$, and  $\bC_{+}$, respectively.
  The constitutive equations are written as
 \begin{gather}\label{siglaw}
	\bsig(\beps,\bx)= (1-H (\bx))\bC_-:(\beps(\bx)-\beps_-^p)+ H (\bx)\bC_+:(\beps(\bx)-\beps_+^p)
\end{gather}
where  $\bsig$ and $\beps=(\nabla \bu)^s$ are the stress and strain tensors, respectively, $\bu$ is the displacement, $\beps_\pm^p$ are the strains in  stress-free states that characterize phases plus and minus.

We  use the following notation  for products of vectors and tensors:  $(\bq\cdot\bk)_i= q_{ij}k_j$, $(\bq\cdot\mathbf{m})_{ij}=q_{ik}m_{kj}$, $\bq:\mathbf{m}=q_{ij}m_{ij}$, $(\mathbf{m}\otimes\mathbf{q})_{ijkl}=m_{ij}q_{kl}$, $(\bA\cdot\bq)_{ijkl}= A_{ijkm}q_{ml}$, $(\bA:\bq)_{ij}= A_{ijkl}q_{kl}$, ${(\bA:\bB)_{ijmn}= A_{ijkl}B_{klmn}}$ where $\mathbf{k}$ is a vector, $\bq$, $\mathbf{m}$ are second rank tensors, and $\bA$, $\bB$ are forth rank tensors,
a summation rule by repeating indices is implied.

Assume that the medium is deformed by a  homogeneous external strain $\beps_0$.
The equilibrium equations, traction and displacement continuity conditions at the interfaces $\Gamma$ and boundary conditions in the case of zero body  forces  are
\begin{gather}
\begin{split}
    \mathbf{x}\in \Omega_- \cup \Omega_+:&  \quad  \nabla\cdot\bsig=0, 
    \label{1}
\\
	\mathbf{x}\in\Gamma:&  \quad [\![\bsig]\!]\cdot\mathbf{n}=0, \quad [\![\mathbf{u}]\!]=0,
\\
	\mathbf{x}\in \partial\Omega:& \quad \bu=\beps_0\cdot\bx.
\end{split}
\end{gather}
Here, we denote a jump across the interface by square brackets:
 ${[\![\,\cdot\,]\!]=(\cdot)_+-(\cdot)_-}$.
The above condition states also that at the interface $\Gamma$
the tangent components of the strain are continuous, and the average stain  in $\Omega$  is $\beps_0$:
\begin{gather}\nonumber
\beps_0
=
\int\limits_{\Omega}\beps(\bx) d\Omega.
\end{gather}

\paragraph{Variational problem}
We consider the minimization problem for the energy of the cell, that is we are looking for such microstructure (characteristic function $H(\bx)$) and strains which  correspond to minimum of the Helmholtz free energy $F$  of the cell. Strictly speaking, this problem does not have a solution, only minimizing sequences \cite{BallJames1987}, which we show below. Correspondingly, the Helmholtz free energy tends to its infimum rather than reaches the minimum. However, below we use both terms.

By \eqref{siglaw}, the Helmholtz free energy density  is written as
\begin{equation}\nonumber
	f_{H}(\bx, \beps)=(1-H (\bx))
f_-(\beps )+H(\bx) f_+(\beps )
\end{equation}
where
\begin{equation}\label{f-density}
f_\pm(\beps )=f_{\pm}^0+w_\pm(\beps), \quad w_\pm(\beps)=\dfrac12(\beps -\beps_\pm^p):\bC_\pm:(\beps -\beps_\pm^p),
\end{equation}
$w_\pm$ are the strain energy densities of a material in  phase states ``$\pm$'', $f_{\pm}^0$ are the free energy densities of the phases in stress-free states (so called chemical energies).

Given average strain $\beps_0$, the optimal structure minimizes the  Helmholtz free energy over all possible microstructures and volume fractions of the phases in the phase transformation. This yields to the variational problem for the minimizers $m_+$, $H$ and  $\beps$:
\begin{equation}\label{energy min form1}
\mathcal{J}_f=\inf\limits_{\mbox{\\ \scriptsize{$0\leq m_+\leq 1$}}} \  \;  \inf\limits_{\int\limits_{\Omega}\!\!H \,d\Omega =\, m_+}~~~ \inf\limits_{\phantom{\int\limits_{V_{\alpha}}}\!\!\!\!\!\!\!\!\bepsscript \in\,\mathcal{E}_0} \int\limits_{\Omega}f_H(\bx,\beps)\, d\Omega
%
%
\end{equation}
where $\mathcal{E}_0$ is the set of strain fields with  an average  $\beps_0$.

We perform the minimization over $f_H(\bx,\beps)$  first.  Since $H$ takes only two values, zero and one, the minimum corresponds to a two-well function $f(\beps)$ as follows:
\begin{equation}\nonumber
\inf_H f_H(\bx, \beps)=f(\beps), \quad f(\beps)=\min_{+,-}\{f_-(\beps),f_+(\beps)\}.
\end{equation}

\paragraph{Minimization method of a two-well energy}

Next, we minimize  \eqref{energy min form1} assuming that  external strains $\beps_0$ and volume fraction $m_+$ are given.  Since   the chemical energies $f^0_{\pm}$  do not make any impact onto the minimization problem at this step,   we focus on finding the  minimum of the strain energy
\begin{equation}\label{energy min form2}
\mathcal{J}_{wm}(m_+, \beps_0)=\inf\limits_{\int\limits_{\Omega}\!\!H \,d\Omega =\, m_+}~~~ \inf\limits_{\phantom{\int\limits_{V_{\alpha}}}\!\!\!\!\!\!\!\!\bepsscript \in\,\mathcal{E}_0}\int\limits_{\Omega}\left((1-H(\bx))w_-(\beps)+H(\bx)w_+(\beps)\right) d\Omega
\end{equation}
This is a problem of optimal composite microstructure.

Finally, the obtained energy can be additionally minimized over $m_+$. This leads to the energy $\mathcal{J}_f(\beps_0)$ of the deformable solid under the phase transition,
\begin{equation}\label{energy min form3}
\mathcal{J}_f(\beps_0)=\inf\limits_{\begin{minipage}{17mm}\scriptsize\centerline{$0 \leq\! m_+ \leq 1$} \end{minipage}} \left\{f^0_-+ m_+\gamma+
\mathcal{J}_{wm} (m_+, \beps_0)\right\}
\end{equation}
where  $\gamma=\sk{f^0}$. Obviously, only the difference $\gamma$ of the chemical energies affects the problem, $f^0_-$ gives just a reference level for the energy calculation. In  phase transition problems the parameter $\gamma$ depends on the temperature and can be treated as an ``energy temperature''.

Notice that the minimization of a two-well Lagrangian \eqref{energy min form2} leads  to optimal solutions with infinitely often oscillating minimizers (strains) that physically corresponds to a microstructured material. 
The cost $\mathcal{J}_{wm}(m_+, \beps_0)$, the infimum over all possible oscillating sequences, is called the relaxed two-well Lagrangian or the quasiconvex envelope of the Lagrangian. The variational problem for $\mathcal{J}_{wm}(m_+, \beps_0)$ possesses a smooth solution that can be found by solving the Euler-Lagrange equation.

The relaxation in \eqref{energy min form2}  is performed by constructing upper and lower bounds for $\mathcal{J}_{wm}(m_+, \beps_0)$,
\begin{equation} \nonumber
\mathcal{PW}(m_+, \beps_0)  \leq \mathcal{J}_{wm}(m_+, \beps_0) \leq \mathcal{LW}(m_+, \beps_0).
\end{equation}
The upper bound $\mathcal{LW}$ is the minimum of the energy  of any microstructure from an a priori chosen class of structures. It is constructed in Section 3.
The lower bound $\mathcal{PW}(m_+, \beps_0) $ used here is the so-called translation bound \cite{Milton1990,Cherkaev2000,Milton2002} that exploits the idea of the polyconvex envelope, see \cite{Dacorogna2008}. The bound is being constructed in Section 4 without direct referring to any specific microstructure and is a structure-independent lower bound for the energy.
Then we demonstrate that the bound is exact by showing that
\begin{equation} \label{exactness-of-bonds}
\mathcal{PW}(m_+, \beps_0)  = \mathcal{LW}(m_+, \beps_0) \quad \forall m_+, \beps_0.
\end{equation}
Thus, we also determine the minimal energy of the composite $\mathcal{J}_{wm} = \mathcal{LW} =\mathcal{PW}$,  solving problem
\eqref{energy min form2}. In other terms, we describe the  quasiconvex envelope of the two-well Lagrangian, see for example \cite{Cherkaev2000}.
In Section 5, we analyze problem \eqref{energy min form3} and construct transformation surfaces.

\section{Laminates with minimal stored energy}
In this section, we calculate the energy $\mathcal{LW}(m_+, \beps_0)$ of simple and higher rank laminates that turn out to be optimal for the considered problem. We  minimize the energy at this class of structures at  given external strains by choosing microstructure parameters and we derive necessary conditions of optimality in terms of the fields inside the  laminates.

\subsection{Lamination formulae}

\begin{figure}[!b]
\begin{center}
\SetLabels
\L (0.55*0.96)  {(a)} \\
\L (0.2*0.96)  {(b)} \\
\L (0.4*0.5) {(c)} \\
\endSetLabels
\begin{center}
\AffixLabels{\centerline{\includegraphics[width=0.6\textwidth]{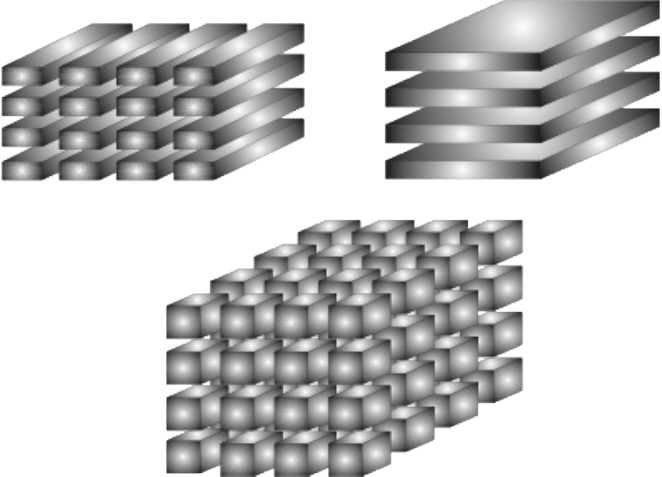}}}
\caption{ Laminated microstructures : (a) simple (first-rank) laminate, (b)  second-rank laminate, (c) third-rank laminate. \label{fig-sloi}}
\end{center}
\end{center}
\end{figure}

\paragraph{Laminates construction}
The rank-n laminate construction \cite{lurie1984exact,Milton1986} is a multistep process. At the first step, one takes a simple
laminate that consists of alternating planar layers occupied by homogeneous phases ``$-$'' and ``$+$''. We call this a simple (or rank-1) laminate.
A second-rank laminate consists of  alternating  layers of the phase ``$-$''  and layers which are themselves simple laminates. The  third-rank laminate consists of the   layers of the phase ``$-$'' and layers which are second-rank laminates (Fig.~\ref{fig-sloi}). The rank-$n$ laminate consists of alternating  layers of the phase ``$-$''   and rank-$(n-1)$ layers.

 In our model, the characteristic size of simple layers is much less than the size of the second-rank layers that in turn is much less than the characteristic  size of the third-rank layers, etc. The strains in such layered material tend to piece-wise constant function when the scale separation increases, as it is  shown in  \cite{Briane1994}.  We will show that the fields in the phase ``$+$" (inclusion) in an optimal laminate belong to the above mentioned PTZ-boundary.

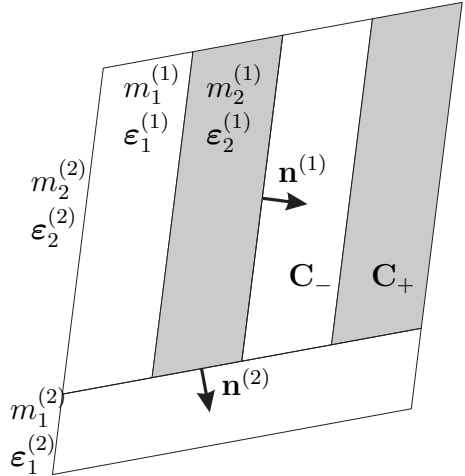
\begin{figure}[!b]
\begin{center}
\SetLabels
\L (0.45*0.8)  $m_{2}^{(1)}$ \\
\L (0.37*0.8)  $m_{1}^{(1)}$ \\
\L (0.28*0.6)  $m_{2}^{(2)}$ \\
\L (0.26*0.12)  $m_{1}^{(2)}$ \\
\L (0.28*0.5)  $\beps_{2}^{(2)}$ \\
\L (0.26*0.02)  $\beps_{1}^{(2)}$ \\
\L (0.52*0.62) $\bn^{(1)}$ \\
\L (0.465*0.17) $\bn^{(2)}$ \\
\L (0.53*0.4) $\bC_-$ \\
\L (0.61*0.4) $\bC_+$ \\
\L (0.37*0.7) $\beps_{1}^{(1)}$ \\
\L (0.45*0.7) $\beps_{2}^{(1)}$ \\
\endSetLabels
\AffixLabels{\centerline{\includegraphics[width=0.4\textwidth]{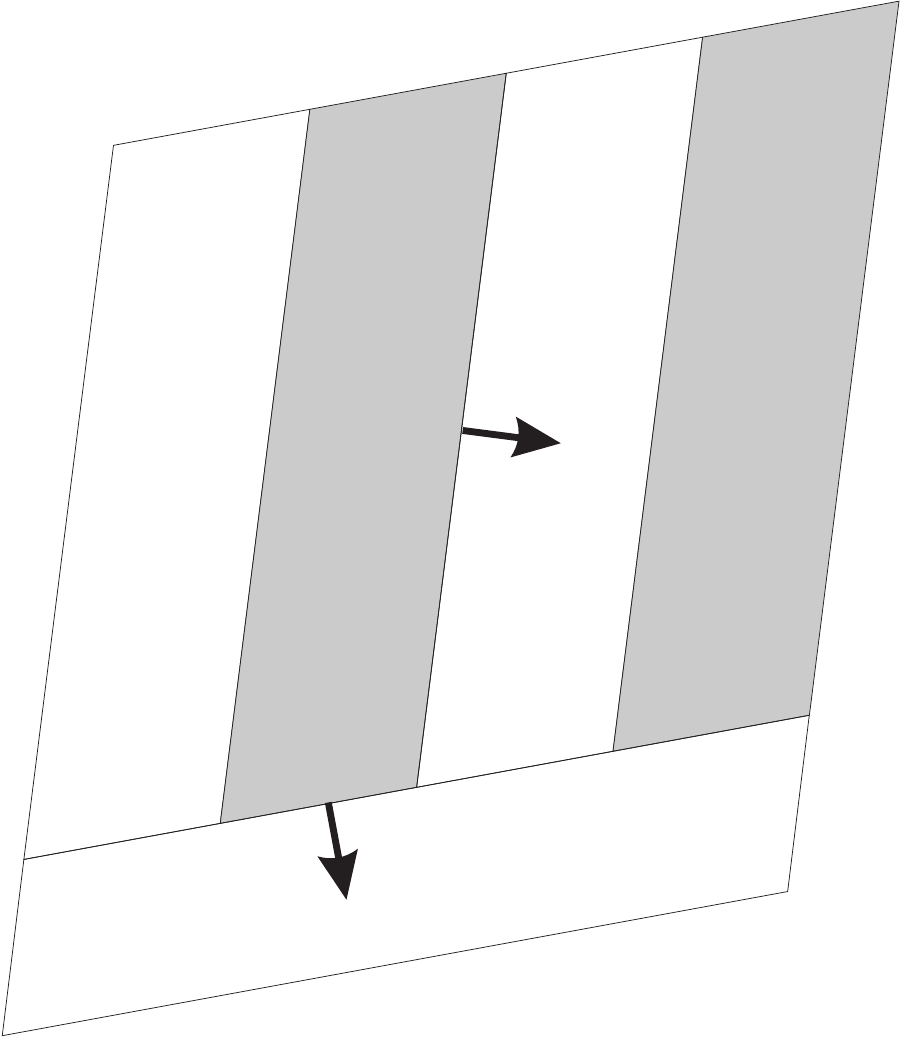}}}
\caption{Second-rank laminate}\label{fig-rank2}
\end{center}
\end{figure}

We use subscripts 1, 2 to identify the ``macroscopic'' layers in laminates. Subscript 1 refers to the layer occupied by the phase ``$-$''. Subscript 2 refers to another layer that is itself a layer of the rank $(k-1)$  in the case of a rank-$k$ laminate. The superscripts ${(1), (2)...}$  in parentheses denote the rank of the laminate. The normals to the layers at various scale levels are $\bn^{(1)}$, $\bn^{(2)}$... (see Fig.~\ref{fig-rank2}).

We denote the volume fraction of the phase  ``$-$''  within the laminate of the rank $k$ as $m_1^{(k)}$, and $m_2^{(k)}$ is the volume fraction of homogenized rank-$(k-1)$ layer, $m_1^{(k)}+m_2^{(k)}=1$.
Let $\beps_1^{(k)}$ be a strain in a phase ``$-$'' layer inside the rank-$k$  layer. Then  $\beps_2^{(k)}$ is a strain averaged within the rank-$(k-1)$ layer.
The strains $\beps_{1,2}^{(k)}$ and stresses  $\bsig_{1,2}^{(k)}$  in each subdomain are uniform, and the following relations are fulfilled for the volume fractions, strains an stresses defined at different scales in the case of the rank-$n$ laminate:
\begin{equation*}\begin{array}{l}
m_1^{(1)}\beps_1^{(1)}+m_2^{(1)}\beps_2^{(1)}=\beps_2^{(2)}, \qquad m_1^{(1)}\bsig_1^{(1)}+m_2^{(1)}\bsig_2^{(1)}=\bsig_2^{(2)},\\
m_1^{(2)}\beps_1^{(2)}+m_2^{(2)}\beps_2^{(2)}=\beps_2^{(3)}, \qquad m_1^{(2)}\bsig_1^{(2)}+m_2^{(2)}\bsig_2^{(2)}=\bsig_2^{(3)},
\\
  ......................................... \\
 m_1^{(n)}\beps_1^{(n)}+m_2^{(n)}\beps_2^{(n)}=\beps_2^{(n+1)}\equiv \beps_0, \quad
  m_1^{(n)}\bsig_1^{(n)}+m_2^{(n)}\bsig_2^{(n)}=\bsig_2^{(n+1)}\equiv \bsig_0,\\
 m_1^{(k)}+m_2^{(k)}=1, \quad k=1,2...n, \quad  \prod\limits_{i=1}^n  m_2^{(i)}=m_+
\\
\bsig_1^{(k)}=\bC_-:(\beps_1^{(k)}-\beps_-^p), \quad
\bsig_2^{(k)}=\bC_2^{(k)}: (\beps_2^{(k)}-\beps_2^{p(k)} )
\end{array}
\end{equation*}
where $\bC_2^{(k)}$ is  the effective elasticity tensor that relates the average stress  and strain  within the rank-$(k-1)$ sublayer of the rank-$k$ laminate, $\beps_2^{p(k)}$ is the effective transformation strain produced by the rank-$(k-1)$ sublayer,  $\bC_2^{(n+1)}\equiv\bC_0$ and $\beps_2^{p(n+1)}\equiv\beps^p_0$ are effective elasticity tensor and effective transformation strain of the composite presented by the rank-n laminate.

To avoid misunderstanding, note that the layer $1$ is always the layer of the phase ``$-$'',  according to the accepted super and subscripts notation $\bC_1^{(k)}=\bC_-$, $\beps_1^{p(k)}=\beps_-^p$  ${(k=1,2...n)}$.
 The strain $\beps_2^{(1)}$ is always a strain inside the domain $\Omega_+$, $\beps_2^{(1)}=\beps_+$.
Further, for the case of the rank-$n$ laminate we express
  $\beps_2^{(1)}=\beps_+$ through $\beps_2^{(2)}$, $\beps_2^{(2)}$ through $\beps_2^{(3)}$,...
 $\beps_2^{(n)}$ through $\beps_0$ and finally derive the dependence of $\beps_+$ on average strain $\beps_0$ and microstructure parameters $\bn^{(k)}$  and $m_1^{(k)}$  ($k=1, 2...n$).

\paragraph{Interaction energy}
The Helmholtz free energy$$ F=\int\limits_\Omega  f_H(\bx,\beps) d\Omega
$$
 of a two-phase body $\Omega=\Omega_-\cup\Omega_+$ at prescribed displacements at the boundary $\partial \Omega$
 can be presented as a sum of the free energy
$$ F_0=\int\limits_\Omega f_-(\beps)d\Omega
$$
of a homogeneous body in a phase state ``$-$'' calculated at the same boundary conditions  and the interaction energy $E$:
\begin{equation} \nonumber
	E=F-F_0.
\end{equation}
The advantage of such a representation is that $F_0$ does not depend on microstructure, and the interaction energy is expressed in a form of an integral over the closed domains $\Omega_+$ occupied by the phase ``$+$'' \cite{Eshelby1957, Mura}. Moreover, as it was mentioned above, the strain is uniform inside $\Omega_+$ in the case of the laminates. This will additionally simplify the consideration.

If the free energy densities of the phases are given by \eqref{f-density} then (see, e.g., \cite{Freidin2007})
\begin{equation*}
E=\int\limits_{\Omega_+}\!\!\left(\gamma+\frac{1}{2}\sk{\beps^p\!:\!\bC\!:\!\beps^p}
+\frac{1}{2} \left(\beps_+\!:\!\sk{\bC}\!:\!
\beps_0-(\beps_++\beps_0)\!:\![\![\bC\!:\!\beps^p]\!]\right)\!\right)d\Omega,
\end{equation*}
where $\gamma=\sk{f^0}$.

Further we will assume that the inverse $\sk{\bC}^{-1}$ exists. Then the expression of the interaction energy can be rewritten in a form \cite{Kublanov1988,Freidin2007}
\begin{equation}\nonumber
	E = \int\limits_{\Omega_+} \left(\gamma_*+\dfrac12\be_0{(\bx)}:\bq_+{(\bx)}\right)d\Omega
\end{equation}
where
\begin{gather}\label{e0def}
\be_0=\beps_0-\sk{\bC}^{-1}:\sk{\bC:\beps^p},\\
\label{qplusdef}
\bq_+(\bx)= \sk{\bC}:\beps_+(\bx) -\sk{\bC:\beps^p},\\
\nonumber
\gamma_*=\gamma +\dfrac12\sk{\beps^p}:\sk{\bC^{-1}}^{-1}:\sk{\beps^p},
\end{gather}
$\beps_0(\bx)$ is the strain that would be in the domain $\Omega_+$ in the case of a homogeneous body in the phase state ``$-$'', $\beps_+(x)$ is the strain inside $\Omega_+$ in the case of the two-phase body.

In the optimal composite problem the volume fractions $m_+$ or $m_-=1-m_+$ of the phases are fixed, and therefore we will temporary exclude from consideration the chemical energies $f_-^0$ and $f_+^0$ and focus on the minimization of the strain energy, i.e. on the calculation  and minimization of
\begin{equation*}
	E_q=\dfrac12\int\limits_{\Omega_+} \be_0:\bq_+ d\Omega
\end{equation*}
for  laminates of various ranks.

\paragraph{Strain energy of laminates}

 The inhomogeneities affect stress-strain state of a body only through the tensor $\bq_+$ \cite{Kunin2,Kanaun2008}, and  we will further
 use  a formula that follows from the displacement and traction continuity conditions \eqref{1} at the interfaces and relates the jump in strain across the interface with the tensor $\bq_+$ and, thus, with the strain $\beps_+$ \cite{Kunin2} (see also \cite{Kanaun2008,FreidinEshelby}).
 If $\beps_\pm$ are the strains at the interface then in a point with a normal~$\bn$
\begin{equation}\label{eq_02}
	\sk{\beps} = -\bK_-(\bn):\bq_+
\end{equation}
where $\bK_-(\bn)=(\bn\otimes\bG_-(\bn)\otimes\bn)^s$, $\bG_-(\bn) = (\bn\cdot\bC_-\cdot\bn)^{-1}$, superscript $s$ means symmetrization:
\begin{equation*}
	K_{ijkl}=n_{(i}G_{j)(k}n_{l)}.
\end{equation*}

Since the tensor $\bq_+$ is uniform inside $\Omega_+$,
\begin{equation}\label{Eq-int}
	E_q=\dfrac12 m_+ \be_0:\bq_+
\end{equation}
Thus, to calculate $E_q$ in the case of the rank-$n$ laminate one has to express $\bq_+$ through $\be_0$. Obviously, $\bq_+$  is a linear function of $\be_0$. It can be derived (see Appendix~A) that the use of shifted strain $\be_0$ instead of $\beps_0$   results in the simplest transmitting formula. For the rank-$n$ laminate
\begin{equation}\label{qplusM}
	\bq_+={\bf M}_-^{(n)}:\be_0
\end{equation}
where  $\be_0$ and $\bq_+$ are defined by \eqref{e0def} and \eqref{qplusdef}, the transmitting tensor
\begin{gather}
\label{Mn}
{\bf M}_-^{(n)}=\left(\sk{\bC}^{-1}+m_- \sum\limits_{i=1}^{n}\beta^{(i)}\bK_-(\bn^{(i)})\right)^{-1},
\end{gather}
microstructure parameters
\begin{equation}\label{beta-def}
\beta^{(i)}=\dfrac{m_1^{(i)}}{m_-m_2^{(i)}}\prod\limits_{j=1}^{i}m_2^{(j)} \quad (i=1,2...n)
\end{equation}
satisfy the restriction
\begin{equation}\label{beta}
\quad \sum\limits_1^n \beta^{(i)} =1.
\end{equation}
Tensors
\begin{gather*}
\bq_2^{(i)}=
 \left(\bC_2^{(i)}-\bC_-\right):\beps_2^{(i)} -\left(\bC_2^{(i)}:\beps_2^{p(i)}-\bC_- :\beps_-^p\right),
 \quad i=1,2...n
 \end{gather*}
 are related with $\bq_+$ as
 \begin{gather}
\label{q2i-qplus}
\bq_2^{(i)}=\dfrac1{m_2^{(2)}}\left(\prod\limits_{i=1}^{i}m_2^{(i)}\right)\bq_+, \quad
\bq_2^{(n)}=\frac{m_+}{m_2^{(n)}}\bq_+.
\end{gather}

Average stress  and strain  within the rank-$i$ sublayer of the rank-$k$ laminate
\begin{gather*}
\bsig_2^{(i+1)}=\bC_2^{(i+1)}:\left(\beps_2^{(i+1)}-\beps_2^{p(i+1)}\right)
\end{gather*}
where the effective elasticity tensors of the sublayer and effective transformation strain
\begin{gather*}
\bC_2^{(i+1)}=\bC_-+m_+\bM_-^{(i)},\\
\beps_2^{p (i+1)}=\beps_-^p+
m_+\left(\bC_2^{(i+1)}\right)^{-1}:\bM_-^{(i)}:\sk{\bC}^{-1}:\bC_+:\sk{\beps^p}.
\end{gather*}

Average stress and strain for the rank-$n$ laminate are related as
\begin{equation}
\label{sigeps-aver}
\bsig_0=\bC_0:\left(\beps_0-\beps_0^{p}\right),
\end{equation}
where effective elasticity tensor and effective transformation strain are, respectively,
\begin{equation}\label{effective}
\bC_0=\bC_-+m_+\bM_-{(n)}, \quad
\beps_0^{p}=\beps_-^p+
m_+\left(\bC_0\right)^{-1}:\bM_-^{(n)}:\sk{\bC}^{-1}:\bC_+:\sk{\beps^p}.
\end{equation}

Finally, substituting \eqref{qplusM} into \eqref{Eq-int} we obtain the formula for the strain energy of the rank-$n$ laminate:
\begin{multline} \label{eq:rank-n energy}
	W^{(n)}=\dfrac12(\beps_0-\beps_-^p):\bC_-:(\beps_0-\beps_-^p)\\ + \dfrac12m_+ \sk{\beps^p}:\sk{\bC^{-1}}^{-1}:\sk{\beps^p}+\dfrac12m_+\be_0:{\bf M}_-^{(n)}:\be_0.
\end{multline}

Of course, by \eqref{sigeps-aver} and \eqref{effective},
$$
W^{(n)}=\dfrac12 \left(\beps_0-\beps_0^{p }\right):\bC_0:\left(\beps_0-\beps_0^{p }\right),
$$
but the representation  \eqref{eq:rank-n energy} demonstrate  explicitly the input of the second phase into the strain energy of the composite in comparison with an one-phase state.

Expressions similar to \eqref{eq:rank-n energy} for the energy of rank-$n$ laminates were earlier derived in matrix form at zero transformation strains  \cite{GibianskyCherkaev1997, Cherkaev2000}, see also \cite{Allaire1997} and reference therein. The  form \eqref{eq:rank-n energy}  takes into account the transformation strains and shows that it is natural to use a shifted strain~$\be_0$ instead of $\beps_0$ in this case. Laminated microstructures have also been considered in applications to phase transformations (see e.g. \cite{Roytburd1998-1,lurie1988certain,Goldsztein2001JMPS,StukiewiczPetryk-JMPS2002} and reference therein). In  Appendix~A, the lamination formulae \eqref{q2i-qplus}--\eqref{eq:rank-n energy} are derived in terms of the notation accepted in the present paper.


The following Proposition relates the sign of the tensor $\bM_-^{(n)}$ with the sign of the jumps $\sk{\bC}$ (see the proof in  Appendix A).
\begin{proposition}\label{prop1}
If the jump $\sk{\bC}$ of the elasticity tensor is a sign-definite tensor then the transmitting tensor $\bM_-^{(n)}$ is of the same sign-definiteness.
\end{proposition}

The proof also shows the existence of $\bM_-^{(n)}$. Additional analysis is needed if the jump $\sk{\bC}$ is not sign-definite.  To confirm  the consistence of the results note that, by \eqref{eq:rank-n energy}, if the transformation strain is zero, i.e. $\sk{\beps^p}=0$, then the strain energy decreases if $\sk{\bC^{-1}}<0$ and increases if $\sk{\bC^{-1}}>0$, as it is to be. 

\paragraph{Relations between strains averaged over the phases}

 Microstructure parameters $\beta_i$  are positive and satisfy the restriction \eqref{beta}.   The sense of $\beta^{(i)}$ is clarified by the formula for the difference of strains averaged within the phases. Let
 \begin{equation}\label{obeps}
 \obeps_\pm=\frac{1}{m_\pm}\int\limits_{\Omega_\pm} \beps(\bx) d\Omega.
 \end{equation}
 Then from the fact that \eqref{qplusM} is  equivalent to  the equation
 \begin{equation*}
 \beps_++m_-\sum\limits_{i=1}^{n}\beta^{(i)}\bK_-(\bn^{(i)}):\bq_+=\beps_0
 \end{equation*}
 and from the equalities
\begin{equation}\label{av_strains}
m_-\obeps_-+ m_+\obeps_+=\obeps_+-m_-(\obeps_+-\obeps_-)=\beps_0, \quad  \obeps_+=\beps_+
\end{equation}
 it follows that
\begin{equation}\label{average-dif}
	\obeps_+-\obeps_-=
-\sum\limits_{i=1}^{n}\beta^{(i)}\bK_-(\bn^{(i)}):\bq_+.
\end{equation}

Formula \eqref{average-dif} allows to formulate strain compatibility conditions in terms of strains averaged over the phases.
Displacement continuity across the interface corresponds to the fulfilling the identity
$$\bK(\bn):\btau\otimes\btau'\equiv 0\quad \forall \btau,\btau'\perp\bn.$$
In the case of a simple laminate $\obeps_\pm=\beps_\pm$, strain compatibility at the interface means that
\begin{gather}\label{compat1}(\obeps_+-\obeps_-):\btau\otimes\btau'=0 \quad \forall \btau,\btau'\perp\bn^{(1)}.\end{gather}

In the less obvious case of the second-rank laminate such that $\bK_-(\bn^{(1)}):\bq_+\neq \bK_-(\bn^{(2)}):\bq_+$   from
 \eqref{average-dif} it follows that
 \begin{gather}\label{compat2}
 (\obeps_+-\obeps_-):\btau\otimes\btau=0 \qquad  \forall\btau\parallel{\bn^{(1)}\times \bn^{(2)}}
 \end{gather}
 The case  $\bK_-(\bn^{(1)}):\bq_+ = \bK_-(\bn^{(2)}):\bq_+$ is not of interest for the minimization problem since can be reduced to the consideration of the simple laminate, as it is shown below.
Of course, \eqref{compat1}, \eqref{compat2} can be directly derived from strain jumps representations  in dyadic forms
\begin{equation*}
\beps_2^{(1)}-\beps_1^{(1)} = (\bn^{(1)}\otimes\ba^{(1)})^s\quad \mbox{and} \quad \beps_2^{(2)}-\beps_1^{(2)} = (\bn^{(2)}\otimes\ba^{(2)})^s,
\end{equation*}
where $\ba^{(1)}$ and $\ba^{(2)}$ are amplitudes of jumps.
The compatibility conditions \eqref{compat1}, \eqref{compat2} will be used in Section~4 as a hint for the choice of the translator parameters.

\subsection{Minimization of the strain energy of  laminates}

The energy of the laminates can be minimized over the normals $\bn^{(i)}$, structural parameters $\beta^{(i)}$, and the rank $n$:
\begin{equation*}
	\mathcal{LW}=\min\limits_{(n)}\min\limits_{\bn^{(i)}}\;\min\limits_{\beta^{(i)}:\beta^{(i)}>0, \sum\limits_{i=1}^{n}\beta^{(i)}=1}\;W^{(n)}.
\end{equation*}

First of all we are
searching  the normals $\bn^{(i)}$ and structural parameters $\beta^{(i)}$ satisfying the conditions
\begin{gather}\label{min_beta}
	\dfrac{\partial \left( W^{(n)}-\zeta(\sum\limits_{i=1}^n \beta^{(i)}-1)\right)}{\partial \beta^{(i)}}=0,\\
\label{min_n}
	\dfrac{\partial\left( W^{(n)}-\sum\limits_{i=1}^n\kappa^{(i)}(\bn^{(i)}\cdot\bn^{(i)}-1)\right)}{\partial \bn^{(i)}}=0
\end{gather}
where $\kappa^{(i)}$ and $\zeta$ are the Lagrange multipliers.

Since  $\delta_{\beta^{(i)}}\bM_-^{(n)}=-m_-\bM_-^{(n)} : \bK_-(\bn^{(i)}): \bM_-^{(n)}\delta\beta^{(i)}$, the condition \eqref{min_beta}
takes the form
\begin{equation}\label{zeta}
	 m_-m_+\bq_+:\bK_-(\bn^{(i)}):\bq_++2\zeta=0, \quad i=1,2...n
\end{equation}
from where it follows that, in the case of the rank-$n$ laminate ($n\geq 2$), the tensor $\bq_+$ and the normals are to be such that the quadratic form has the same values for all $\bn^{(i)}$ $(i=1,2...n)$.

Since  $\delta_{\bn^{(i)}}\bM_-^{(n)}=-\beta^{(i)}m_-\bM_-^{(n)} : \delta_{\bn^{(i)}}\bK_-(\bn^{(i)}): \bM_-^{(n)}$, the  condition \eqref{min_n} can be rewritten as
\begin{equation}\label{qKq_var}
	\dfrac12m_-m_+\beta^{(i)}\bq_+:\delta \bK_-(\bn^{(i)}):\bq_+ +\kappa^{(i)}\delta\bn^{(i)}\cdot \bn^{(i)}=0
\end{equation}
where \eqref{qplusM} is accounted for.

Eqs. \eqref{qKq_var} look like the extremum conditions of the quadratic forms
$$
\mathcal{K}_-(\bn^{(i)} \,|\, \bq_+)=\bq_+:\bK_-(\bn^{(i)}):\bq_+,\quad \bn^{(i)}\cdot\bn^{(i)}=1
$$
at given $\bq_+$, and the maximum of $\mathcal{K}_-$ corresponds to minimum of $W^{(n)}$.
Maximal and minimal values of $\mathcal{K}(\bn)$ have been found in \cite{MorozovFreidin1998,FreidinUniv1999} in the context of the PTZ construction. The maximal value corresponds to external PTZ boundary related, as it was mentioned above, with locally stable interfaces. The maximal value  also corresponds  to the minimal energy of the simple laminates \cite{FreidinSharipova2006}.

If   the phases are isotropic then
\begin{gather}\nonumber
	\bC_\mp=\lambda_\mp\bE\otimes\bE+2\mu_\mp\bI, 
\\ \nonumber
\bK_-(\bn)=\dfrac1{\mu_-}\left((\bn\otimes\bE\otimes\bn)^s- \dfrac1{2(1-\nu_-)}\bn\otimes\bn\otimes\bn\otimes\bn\right)
\end{gather}
and therefore
\begin{gather}
\label{i5}
\bK_-(\bn)\!:\!\bq =\frac{1}{2\mu_-}\left(\bq_n\otimes
\bn + \bn \otimes \bq_n -
\frac{q_{nn}}{1-\nu_-}\bn\otimes\bn\right),\quad
\\
\label{NN}
\mathcal{K}_-(\mathbf{n},\mathbf{q})=\frac{1}{\mu_-}( N_{2}-\varkappa_- N_{1}^{2}),
 \quad \varkappa_-=\frac{1}{2(1-\nu_-)}
\end{gather}
Here $\bE$ and $\bI$ are the second and forth rank unit tensors, in Cartesian coordinates $E_{ij}=\delta_{ij}$, $I_{ijkl}=\dfrac{1}{2}(\delta_{ik}\delta_{jl}+\delta_{il}\delta_{jk})$, $\delta_{ij}$ is Kronecker's delta,  $\lambda_\mp$ and $\mu_\mp$ are the Lame coefficients, $\nu_-$ is Poisson's ratio, $\bq_n=\bq\cdot\bn$, \
$N_{1} = q_{nn}=\bn\cdot \bq\cdot \bn$ and $N_{2} = \bn \cdot \bq^2\cdot\bn $
are so-called orientation invariants \cite{MorozovFreidin1998,FreidinUniv1999},   $q_i$ ($i=1,2,3$) are the eigenvalues of  $\bq$.

It was shown \cite{MorozovFreidin1998,FreidinUniv1999} that
the direction of the normal providing the maximum
$$
\mathcal{K}_-^{max}=\max\limits_{\bn:\; |\bn|=1}\mathcal{K}_-(\bn\,|\,\bq_+ )
$$
depends on the relationships between eigenvalues of $\bq_+ $ as follows.

Assume that the tensor $\bq_+$ is given and has different eigenvalues, and $q_{\min}$, $q_{\max}$ and $q_{\rm mid}$, and $\be_{max}$, $\be_{min}$, $\be_{mid}$ are the minimal,
maximal and intermediate eigenvalues and corresponding eigenvectors of $\bq_+$,  $n_{min}$, $n_{\rm max}$ and
$n_{\rm mid}$ are  components of the maximizing normal $\bn^\ast$
in the basis of eigenvectors of $\bq_+$,
$|q|_{\max}$ is the  maximal absolute
eigenvalue  and $\be_{|q|{\max}}$
is the corresponding eigenvector of $\bq_+$, $|q|_{\min}$ is the  minimal absolute
eigenvalue of $\bq_+$. Then if
\begin{gather}\label{normal-1}
	q_{min}q_{max}<0 \quad \textrm{or} \quad \left\{
	\begin{array}{l}
		q_{min}q_{max}>0\\
		(1-\nu_-)|q|_{min}<\nu_- |q|_{max}
	\end{array}\right.
\end{gather}
then the normal lies in the plane of maximal and minimal eigenvalues of the tensor $\bq_+$, $n_{mid}=0$, and the components are
\begin{gather}\label{normal}
	n_{max}^2=\dfrac{(1-\nu_-)q_{max}-\nu_-q_{min}}{q_{max}-q_{min}}, \quad
n_{min}^2=\dfrac{\nu_-q_{max}-(1-\nu_-)q_{min}}{q_{max}-q_{min}}.
\end{gather}
In this case, by \eqref{i5} and \eqref{NN},
\begin{gather}
\label{Kq-dif}
	\bK_-(\bn^\ast):\bq_+= \dfrac{q_{max}-q_{min}}{2\mu_-}
(n^2_{max}\be_{max}\otimes\be_{max} - n^2_{min}\be_{min}\otimes\be_{min} ),
\end{gather}
\begin{multline}
\label{Kq-dif1}
\mathcal{K}_-^{max}=\dfrac{q_{max}-q_{min}}{2\mu_-}
(n^2_{max}q_{max}  - n^2_{min}q_{min} )\qquad\qquad\qquad\qquad\qquad
\\
\qquad\qquad=\dfrac{1-\nu_-}{2\mu_-}(q_{max}^2+q_{min}^2)-\dfrac{\nu_-} {\mu_-}q_{max}q_{min}.
\end{multline}

Otherwise
\begin{gather}\label{normal-2}
	\bn^\ast =\be_{|q|max}, \\ \label{Kq-dif_dirrect}
\bK_-(\bn^\ast):\bq_+= \dfrac{1-2\nu_-}{2\mu_-(1-\nu_-)}|q|_{max}\be_{|q|max}\otimes\be_{|q|max}, \\
\label{Kmax}
\mathcal{K}_-^{max}=\dfrac{1-2\nu_-}{2\mu_-(1-\nu_-)}|q|_{max}^2.
\end{gather}

Let $\bq_+$ be axisymmetric: $\bq_+=q(\bE-\bk\otimes\bk)+q_k\bk\otimes\bk$. If
\begin{gather}\label{qaxi1}
	q_k q<0 \quad \textrm{or} \quad \left\{
	\begin{array}{l}
		q_kq >0\\
		(1-\nu_-)|q|_{min}<\nu_- |q|_{max}
	\end{array}\right.\\
\nonumber |q|_{min}=\min \{|q_k|,|q|\},\quad |q|_{max}=\max \{|q_k|,|q|\}
\end{gather}
then maximum of $\mathcal{K}_-$ is reached at any normal with the projection onto the axe $\bk$ \begin{gather}\label{qaxi2}
	n_{k}^2=\dfrac{(1-\nu_-)q_{k}-\nu_-q}{q_k-q}.\end{gather}
Then
\begin{gather}
\label{qaxi2-1}
\bK_-(\bn^\ast):\bq_+=\frac{q_k-q}{2\mu_-} \left(n_k^2\bk\otimes\bk-(1-n_k^2)\be \otimes\be \right),
\\
\label{qaxi2-3}	\mathcal{K}_-^{max}=\dfrac{q_k-q}{2\mu_-}\left(q_kn_k^2-q(1-n_k^2)\right)=\dfrac{1-\nu_-}{2\mu_-}(q_k^2+q^2)-\dfrac{\nu_-}{\mu_-}q_kq,
\end{gather}
where
$\be =\bk\times\dfrac{\bk\times\bn^\ast}{|\bk\times\bn^\ast|}$ is the eigenvector of $\bK_-(\bn^\ast):\bq_+$ lying in the plane containing the vectors~$\bk$ and~$\bn^\ast$.

Otherwise
\begin{gather} \label{qaxi3}
		\bn^\ast =\bk \ \mbox{if} \  |q_k|>|q|,\\
\label{qaxi4}
		\bn^\ast \perp\bk \ \mbox{if} \  |q_k|<|q|,\\
\label{qaxi5}
\bK_-(\bn^\ast):\bq_+= \dfrac{1-2\nu_-}{2\mu_-(1-\nu_-)}\max \{|q_k|,|q|\})\bn^\ast \otimes\bn^{\ast}, \\
\label{Kmax2}
\mathcal{K}_-^{max}=\dfrac{1-2\nu_-}{2\mu_-(1-\nu_-)}(\max \{|q_k|,|q|\})^2.
		\end{gather}

If $\bq_+=q\bE$ then
\begin{equation}\label{KqE}\bK_-(\bn ):\bq_+ =cq\bn\otimes\bn,
\end{equation}
$\mathcal{K}_-$ does not depend on $\bn$,
\begin{gather}\label{cK}
\bq_+ :\bK_-(\bn ):\bq_+ =cq^2, \quad
c=\frac{1-2\nu_-}{2\mu_-(1-\nu_-)}=
\left(k_-+\frac{4}{3}\mu_-\right)^{-1},
\end{gather}
$k_-$ is the bulk modulus.

If the normal $\bn^\ast$ corresponds to the maximum of  $\mathcal{K}_-(\mathbf{n},\mathbf{q}_+)$ then the jump of strains has the same eigenvectors as the tensor $\bq_+$. Thus, since the both phases are isotropic and strains $\beps_\pm^p$ are spherical, and the strain tensors $\beps_+$ and $\beps_-$ at the interfaces satisfying the relationships \eqref{normal-1}, \eqref{normal} or \eqref{normal-2} or \eqref{qaxi1}--\eqref{qaxi4} have the same eigenvalues \cite{MorozovFreidin1998}. Note that  from \eqref{i5} one can
 see that $\bq_+=\sum q_i\be_i\otimes\be_i$ and $\bK_-(\bn):\bq_+$ have the same eigenvectors $\be_i$ ${(i=1,2,3)}$ if and only if
\begin{gather}\label{eKe}
\be_i\cdot\left(\bK_-(\bn):\bq_+\right)\cdot{\be}_j =\frac{1}{2\mu_-}\left(q_i+q_j-\frac{q_{nn}}{1-\nu_-}\right)n_in_j=
 \left\{
	\begin{array}{ll}
		0 &  \mbox{if} \ \; i\neq j,\\
		K_i &  \mbox{if} \ \;  i= j.
	\end{array}\right.
\end{gather}
where $K_i$ are the eigenvalues of $\bK_-(\bn):\bq_+$. All the cases listed above are particular cases of \eqref{eKe}.

Taking into account relationships \eqref{NN}--\eqref{cK},  we can \emph{a priory} characterize the strains in the phase  ``$+$'' and the directions of the normals in the laminates of various ranks which can satisfy both conditions \eqref{qKq_var} and \eqref{zeta}. To distinguish different cases of interest we will use the following properties of strains in laminates (see derivations and proofs in Appendix B):

1. Strains at different scales are related by a transmitting formula
\begin{gather}\label{eq_eps_dif}
\beps_1^{(i)}-\beps_1^{(i-1)}= \left(\prod\limits_{j=1}^{i-1}m_2^{(j)}\right)\left(\bK_-(\bn^{(i)})-\bK_-(\bn^{(i-1)})\right):\bq_+.
\end{gather}

2. We formulate another feature as
\begin{proposition}\label{permute} Let two rank-$n$ laminates $A$ and $B$ differ by the permutation of adjacent normals $\bn^{(k)}$ and $\bn^{(k+1)}$ (${1\leq k\leq n-1}$) such that
\begin{equation*}
	\begin{array}{l}
		\bn_B^{(i)}=\bn_A^{(i)} \quad \mbox{if \; $1\leq i<k-1$ \; and \; $k+1< i\leq n$   }\\
		\bn_B^{(k)}=\bn_A^{(k+1)},\quad
		\bn_B^{(k+1)}=\bn_A^{(k)}
	\end{array}
\end{equation*}
and the volume fractions
$m_{1A,B }^{(k)}$,    $m_{1A,B }^{(k+1)}$ and $m_{2A,B }^{(k)}$
obey the relationships
\begin{equation}\label{mAB}
m_{1A }^{(k)}=m_{1B}^{(k+1)}m_{2B}^{(k)}, \quad m_{1B }^{(k)}=m_{1A}^{(k+1)}m_{2A}^{(k)}.
\end{equation}
Then, given average strain $\beps_0$, the laminates have the same strain energy.
\end{proposition}

The above statements  are used in the proof of the following proposition that allows to reduce the rank of the energy minimizing laminate without changing its energy.

\begin{proposition}\label{prop2}  If two normals in a laminate satisfy the equality
\begin{equation}\label{KqCorol}
\bK(\bn^{(i)}):\bq_+=\bK(\bn^{(j)}):\bq_+
\end{equation}
then the rank of the laminate can be lowered by a unit without
changing the strain energy.
\end{proposition}

Further, for various combinations of eigenvalues of $\bq_+$, we will choose the optimal laminates of the lowest rank among energy equivalent laminates of the first, second and third rank. It can be proved that
there are only four cases which represent all energy nonequivalent laminates which  satisfy the conditions \eqref{min_n} and \eqref{min_beta} (see Appendix C). These cases are:

I. \emph{Simple (first-rank) laminates}.
The laminate  provides the maximum of $\mathcal{K_-(\bn, \bq_+)}$ and
has the normal determined by \eqref{normal} either \eqref{normal-2}, depending on the eigenvalues of $\bq_+$.

II. \emph{Second-rank laminates}. The tensor $\bq_+$ is axisymmetric, $$\bq_+=q(\bE-\bk\otimes\bk)+q_k\bk\otimes\bk$$

\begin{itemize}
\item[(a)] The eigenvalues $q$ and $q_k$ satisfy \eqref{qaxi1}.
     Then any two different normals with the same square of the projection onto the axe $\bk$ equal to \eqref{qaxi2}  and noncoplanar with $\bk$ can be taken as $\bn^{(1)}$ and~$\bn^{(2)}$.

     \item[(b)] The eigenvalues $q$ and $q_k$ do not satisfy \eqref{qaxi1} and $|q_k|<|q|$. Then two different normals $\bn^{(1)}$ and  $\bn^{(2)}$ lie  in a plane perpendicular to the
         axe~$\bk$.
     \end{itemize}

     III.  \emph{Third-Rank laminates}. The tensor $\bq_+$ is spherical, any  three noncoplanar unit vectors can be the normals.

     It can be also  proved that optimal laminates of the ranks higher than three are energy equivalent to one of the above-mentioned optimal laminates of rank-1,2 or 3 (see Appendix C).

     Note that the second-rank laminates are of interest only with axisymmetric tensor  $\bq_+$, and the third-rank laminates only with a spherical $\bq_+$. The cases with other $\bq_+$ are reduced to lower rank laminates. Note also that in the case II(a) the normals $\bn^{(1)}$ and  $\bn^{(2)}$ do not lie in the same eigenplane of the tensor $\bq_+$, as opposed to II(b).

\section{Translation  bound for  strain energy}

\subsection{Translated energy}

\paragraph{Translator} We construct the lower bound of the optimal energy using the translation method \cite{Cherkaev2000,GibianskyCherkaev1997,lurie1983optimal,Milton1990, Milton2002} that is based on the polyconvexity \cite{Dacorogna2008}. The construction follows
 \cite{Cherkaev2000, GibianskyCherkaev1997} (see also \cite{Bhattacharya2008}).

 According to the translation method, we replace differential properties of a strain
 $ \beps =(\nabla \bu)^s$
 with weaker integral corollaries  of quasiconvexity: any subdeterminant of Jacobian $\nabla \bu$ is quasiafine:   $\langle$subdet$(\nabla \bu)\rangle=$subdet$\langle\nabla \bu\rangle$ where  $\langle\,\cdot\,\rangle$ denotes averaging over $\Omega$ (see, e.g., \cite{reshetnyak1967general}).

  The nonconvex function  $\theta_1=\beps_{23}^2-\beps_{22}\beps_{33}$ can be presented as
\begin{equation*}  
\theta_1= \left(\dfrac{\partial u_2}{\partial x_3} \dfrac{\partial u_3}{\partial x_2} - \dfrac{\partial u_2}{\partial x_2} \dfrac{\partial u_3}{\partial x_3}\right) + \frac{1}{4} \left(\dfrac{\partial u_2}{\partial x_3} - \dfrac{\partial u_3}{\partial x_2}\right)^2,
\end{equation*}
  the sum of such subdeterminant and a convex function, where $u_i$ are the components of the displacement vector $\bu$. Therefore,   Jensen's inequality is satisfied
 \begin{equation} \nonumber 
   \theta_i \left(\beps_0\right) \leq \int_\Omega \theta_i(\beps) d\Omega
 , ~~i=1, 2, 3, \quad \beps_0=\int_\Omega \beps(\bx)\, d\Omega
\end{equation}
 where the functions $\theta_2$ and $\theta_3$ are obtained from $\theta_1$  by permutation.
 Multiplying them by three nonnegative numbers $t_1, ~t_2, ~t_3$ and adding the left- and righthand sides, we obtain the inequality of a translator:
\begin{equation} \label{jensen}
\varphi\left(\bt, \beps_0\right) \leq \int_\Omega \varphi(\bt,\beps) \, d\Omega   ~~
\forall t_i\geq 0, ~i=1, 2,3
\end{equation}
where translator $\varphi(\bt,\beps)$  (see \cite{GibianskyCherkaev1997})
\begin{equation}\label{translator}
\varphi(\beps, t_1, t_2, t_3)=t_3(\varepsilon_{12}^2-\varepsilon_{11}\varepsilon_{22}) + t_2(\varepsilon_{13}^2-\varepsilon_{11}\varepsilon_{33}) + t_1(\varepsilon_{23}^2-\varepsilon_{22}\varepsilon_{33})
\end{equation}
is a function dependent of three nonnegative parameters $t_1, ~t_2$ and $t_3$.

The quadratic form of the translator can be written in tensorial form
 $\varphi(\bt,\beps)= \dfrac12 {\beps}:\bT :{\beps}$, where
\begin{equation}
\label{T}
	\bT =(\bt\otimes\bE+\bE\otimes\bt)+(\tr\,\bt)(\bI-\bE\otimes\bE)-(\bI\cdot\bt+\bt\cdot\bI).
\end{equation}
is the corresponding fourth-rank tensor that possesses all symmetries with respect to indices permutations as the elasticity tensor, where tensor $\bt$ is the second-rank tensor with the eigenvalues $t_1$, $t_2$ and $t_3$ (see Appendix~D for derivation of \eqref{T}). Tensorial form is used for the further analysis.

\paragraph{Translation bound} The energy of the composite cell
$$w_0(\beps_0)= \dfrac12(\beps_0-\beps_0^p): \bC_0 :(\beps_0-\beps_0^p), $$
where $\bC_0$ and $\beps_0^p$ are the effective elasticity tensor and the effective transformation strains,  satisfies the variational condition
\begin{equation} \nonumber 
w_0(\beps_0)= \dfrac12 \min\limits_{
\substack{
\beps: \langle\beps \rangle  = \beps_0
\\
\beps=(\nabla\bu)^s
}
} \int_\Omega (\beps(\bx)-\beps^p(\bx)): \bC(\bx) : (\beps(\bx)-\beps^p(\bx))\, d\Omega.
\end{equation}
Subtracting (\ref{jensen}) from both parts of this equality we obtain an inequality 
\begin{equation}\label{wo-transl}
w_0(\beps_0) -\varphi(\beps_0,\bt)\geq \min\limits_{
\substack{
\beps: \langle\beps \rangle  = \beps_0
\\
\beps=(\nabla\bu)^s
}
}\int\limits_{\Omega}\left((1-H(\bx))w_-^t(\beps,\bt)+H(\bx)w_+^t(\beps,\bt)\right)d\Omega, ~~\forall \bt
\end{equation}
where
\begin{equation}\label{wtT}
w^t_\pm(\bt,\beps)=\dfrac12(\beps-\beps^p_\pm):\bC_\pm:(\beps-\beps_\pm^p)-\dfrac12\beps:\bT:\beps
\end{equation}
are the translated strain energy densities.

Assume that $\bT$ is so chosen, that  $w^t_\pm(\bt,\beps)$ is a convex function of $\beps$:
$$\beps:(\bC_\pm -\bT):\beps \geq 0, \quad \forall \beps. $$
 The inequality \eqref{wo-transl} becomes stronger if  we increase the set of minimizers $\beps(x)$ by neglecting differential property $\beps(x)=(\nabla \bu)^s$. Due to convexity of $w^t_\pm(\bt,\beps)$, we  can replace  $\beps(x)$ in each phase by its average $\overline{\beps}_\pm$, respectively.
\begin{equation}\label{wo-transl-2}
w_0(\beps_0) -\varphi(\beps_0,\bt)\geq \min\limits_{
\substack{
m_-\obeps_-+m_+\obeps_+=\beps_0
}
}(m_-w^t_-(\bt,\obeps_-)+m_+w^t_+(\bt,\obeps_+))
~~\forall \bt
\end{equation}
where
$$
\obeps_-=\frac{1}{m_-}\int\limits_{\Omega_-}\beps d\Omega, ~~ \obeps_+=\frac{1}{m_+}\int\limits_{\Omega_+}\beps d\Omega.
$$

Finally, we maximize $\varphi(\beps_0,\bt)$ and the right hand side of
the inequality over $\bt$.
Then the inequality \eqref{wo-transl-2} becomes
\begin{equation} \nonumber
w_0(\beps_0) \geq {\mathcal{PW}}.
\end{equation}
Here,
\begin{equation}\label{defPW}
{\mathcal{PW}}=\max\limits_{t_i\in\Theta} \left(\min\limits_{m_-\obeps_-+m_+\obeps_+= \beps_0}{W^t}(\obeps_-,\obeps_+)+\varphi(\bt,\beps_0)\right)
\end{equation}
and
\begin{equation} \nonumber
	{W^t}(\obeps_-,\obeps_+)=m_-w_-^t(\obeps_-)+m_-w_+^t(\obeps_+),
\end{equation}
the average strains $\obeps_\pm$ are defined by \eqref{obeps} and satisfy the restriction \eqref{av_strains}, and the set $\Theta$ of admissible translator parameters is determined  in Subsection~\ref{t-admissible}.
The function ${\mathcal{PW}}$ is called the translation bound.

\subsection{Translation-optimal strains in the phases}\label{eps-optimal}
Here we find minimizing average strains $\obeps_\pm$. Optimal (maximizing) translator parameters are defined in Subsection~4.3.
Since the function $W^t$ is a convex function of  strains $\obeps_\pm$, the minimization with respect to $\obeps_\pm$ is carried out  by the direct differentiating   ${W^t}$ at fixed  $\beps_0$. Taking into account \eqref{wtT}, we obtain
\begin{equation*}
	\dfrac{\partial }{\partial \obeps_\pm} \left({W^t}+\blambda:(m_-\obeps_-+m_+\obeps_+-\beps_0)\right)=m_\pm\left(\bC_\pm:(\obeps_\pm-\beps_\pm^p)  -\bT:\obeps_\pm+ \blambda \right)=0
\end{equation*}
where $\blambda$ is the second order tensor of Lagrange multipliers by the restriction \eqref{av_strains}. Minimizers $\obeps_-$ and $\obeps_+$ are to be found from the system of equations
\begin{gather*}
\left\{\begin{array}{l}
 (\bC_+-\bT):\obeps_+  -\bC_+:\beps_+^p=  (\bC_- - \bT):\obeps_-  -\bC_-  :\beps_-^p  \\
 m_-\obeps_-+m_+\obeps_+=\beps_0
\end{array}\right.
\end{gather*}
They are:
\begin{gather} \label{average optimal strains}
\begin{split}	
	&\obeps_-=(m_-\bC_++m_+\bC_--\bT)^{-1}:((\bC_+-\bT):\beps_0-m_+\sk{\bC:\beps^p}),\\
	&\obeps_+=(m_-\bC_++m_+\bC_--\bT)^{-1}:((\bC_--\bT):\beps_0+m_-\sk{\bC:\beps^p}),
\end{split}
\end{gather}

\subsection{Domain of admissible  translator parameters}\label{t-admissible}

The translation parameters vary in a set defined by several constraints, see \cite{GibianskyCherkaev1997,Bhattacharya2008}:
(i) by definition of the translator \eqref{translator}, the translator parameters are non-negative, and (ii) they are restricted by the requirement of  {convexity }of $w^t$, that is by nonnegativity of the quadratic forms
\begin{equation}\label{quadratic}
	\begin{split}
		\beps:(\bC_\pm-\bT):\beps=&(\lambda_\pm+2\mu_\pm)(\varepsilon^2_{11}+\varepsilon^2_{22}+\varepsilon^2_{33})+\\
		&2(\lambda_\pm+t_3)\varepsilon_{11}\varepsilon_{22} +
		2(\lambda_\pm+t_2)\varepsilon_{11}\varepsilon_{33} +
		2(\lambda_\pm+t_1)\varepsilon_{22}\varepsilon_{33} + \\
		&2(2\mu_\pm-t_3)\varepsilon_{12}^2 +
		2(2\mu_\pm-t_2)\varepsilon_{13}^2 +
		2(2\mu_\pm-t_1)\varepsilon_{23}^2.
	\end{split}
\end{equation}

The Hessian of  \eqref{quadratic} with respect of entries of $\beps$ is
\begin{equation} \label{Hessian}
	H=2\left(
	\begin{array}{cccccc}
		\lambda_\pm+2\mu_\pm & \lambda_\pm+t_3 & \lambda_\pm+t_2 & 0 & 0 & 0\\
		\lambda_\pm+t_3 & \lambda_\pm+2\mu_\pm &  \lambda_\pm+t_1 & 0 & 0 & 0\\
		\lambda_\pm+t_2 & \lambda_\pm+t_1 & \lambda_\pm+2\mu_\pm & 0 & 0 & 0\\
		0 & 0 & 0 & 2(2\mu_\pm-t_3) & 0 & 0\\
		0 & 0 & 0 & 0 & 2(2\mu_\pm-t_2) & 0\\
		0 & 0 & 0 & 0 & 0 & 2(2\mu_\pm-t_1)
	\end{array}
	\right)
\end{equation}
It is non-negative-definite if
\begin{gather}\label{eq_17_1}
	|t_i'| \leq 1,  \quad i=1,2,3\\ \label{eq_18_1}
	2t'_1t'_2t'_3-\left((t'_1)^2+(t'_2)^2+(t'_3)^2 \right) +1 \geq 0,
\end{gather}
where $t'_i=\dfrac{\lambda_\pm+t_i}{\lambda_\pm+2\mu_\pm}$.

Inequality \eqref{eq_18_1} shows that if one of the translator parameters $t_i'$ equals unity then other two parameters must be equal to each other.
Positivity of translator parameters together with  equation \eqref{eq_17_1} show that the admissible translator parameters are in the range
\begin{equation}\nonumber
 \Theta=\left\{t_i: \ t_i\in[0;2\mu], \ i=\overline{1,3}, \ \mu=\min\{\mu_-,\mu_+\}, \ \text{and inequality \eqref{eq_18_1} holds}\right\}.
\end{equation}
Recall that we assumed $\mu_-<\mu_+$ in laminates construction,  further we follow this assumption.

\subsection{The optimal laminates and the choice of  translator parameters}

Now we have to find optimal translator parameters $t_1,t_2,t_3\in  \Theta$ to compute~\eqref{defPW}.
We do not find translator parameters by straight maximization in \eqref{defPW}. Instead, following approach
suggested in \cite{gibiansky1987microstructures,GibianskyCherkaev1997} and used in \cite{albin2007multiphase,cherkaev2013three}, we assume that the lower bound is exact and attained by laminates which can be referred to as trial microstructures at this stage. We use compatibility conditions \eqref{compat1} and \eqref{compat2} and the properties of the strains $\beps_+$ in optimal laminates to find translator parameters in various domains of strain space. The attainability of
the lower bound with the use of optimal laminates is proved in the next subsection.

Depending on $\beps_0$,
the translator parameters may lie inside the domain $\Theta$ or on various parts of its boundary.
On one hand, the optimal translator parameters maximize the translated energy. On the other hand we expect
that the lower bound is attainable by one of the optimal laminate microstructure. Then,
by the laminates analysis of Section~3,
the translation-optimal average strain
$\obeps_+$ in \eqref{average optimal strains} is spherical or the
difference of the translation-optimal average strains
\begin{gather}	\label{eps_difference}
	\Delta\obeps=\obeps_+ - \obeps_- = (m_-\bC_++m_+\bC_--\bT)^{-1}:((\bC_--\bC_+):\beps_0+\sk{\bC:\beps^p})
\end{gather}
 satisfies compatibility conditions \eqref{compat1} or \eqref{compat2}.
Since the condition \eqref{compat2} corresponds to the second-rank laminates, an additional
restriction on the translator parameters is that the tensor $\obeps_+$  is axisymmetric in this case.
Different domains of external strains correspond to different optimal laminates and, thus, to different strains $\obeps_\pm$. This, in accordance with \eqref{average optimal strains}, gives different sets of the translator parameters.

The difference \eqref{eps_difference} can be rewritten in the form
 \begin{gather}
 	\label{strain_difference}
 	\Delta\obeps=\obeps_+ - \obeps_- = -(m_-\bC_++m_+\bC_--\bT)^{-1}:\bq_0,
 \end{gather}
 where $\bq_0=\sk{\bC}:\beps_0-\sk{\bC:\beps^p}=\sk{\bC}:\be_0$.
 Then it is convenient to present the results through the eigenvalues  $q_i \ (i=1,2,3)$ of   $\bq_0$ instead of $\beps_0$.
 We assume that tensors $\bt$ and $\bq_0$ have the same principal directions
 $\be_1$, $\be_2$ and $\be_3$. Then, by   \eqref{average optimal strains} and \eqref{strain_difference}, the  strain tensors $\obeps_\pm$ and the  difference  $\Delta\obeps$ have the same principal directions.
  This simplifies  finding $t_1,~t_2,~t_3$.

The  calculation of the translator parameters leads
to the following results.
Let  $q_i$ be ordered such that $|q_3|\geq|q_1|,|q_2|$. Then all possible microstructures and corresponding translator parameters can be shown on the plane $\delta_1^q, \delta_2^q$, where $\delta_1^q=q_1/q_3$, $\delta_2^q=q_2/q_3$. Note that $|\delta_i^q|\leq 1$.

The top half $\delta_1^q\leq\delta_2^q$ of square $|\delta_i^q|\leq 1$ is divided into domains $\Psi_i^q$ $(i=1,2...7)$, and each domain is characterized by some laminate microstructure
and by a set of values of the translator parameters:
\begin{gather} \label{optimal translators}
\begin{split}
 t_1=t_2=t_3=2\mu_-   \quad \mbox{ if } \bq_0 \in \Psi_1^q, \\
  t_1=2\mu_-, \quad t_2=t_3=\dfrac{2\delta_1^q(\mu_- +\bar\mu )+\bar\lambda(2\delta_1^q-\delta_2^q-1)}{1+\delta_2^q}    \quad \mbox{  if  } \bq_0 \in \Psi_2^q,\\
t_1=2(\delta_2^q-1)\bar\lambda+2\delta_2^q\bar\mu, \quad t_2=t_3=(\delta_1^q-1)\bar\lambda+2\delta_1^q\bar\mu   \quad \mbox{  if  } \bq_0 \in \Psi_3^q, \\
t_1=2\mu_-, \quad t_2=t_3=0  \quad \mbox{ if } \bq_0 \in \Psi_4^q,\\
t_1=\dfrac{2\bar\mu(\bar\lambda(\delta_1^q-2\delta_2^q+1)- 2\delta_2^q\bar\mu)}{\bar\lambda(1-\delta_1^q)-2\bar\mu}, \quad t_2=t_3=0   \quad \mbox{  if  } \bq_0 \in \Psi_5^q, \\
t_1=t_2=0, \quad t_3=\dfrac{2\bar\mu(2\delta_2^q(\bar\lambda+\bar\mu)- \bar\lambda(1+\delta_1^q))}{\bar\lambda(1-\delta_1^q)+2\bar\mu\delta_1^q}  \quad \mbox{  if  } \bq_0 \in \Psi_6^q, \\
t_1=t_2=0, \quad t_3=2\mu_-  \quad \mbox{ if } \bq_0 \in \Psi_7^q,
\end{split}
\end{gather}
where $\bar\mu=m_+\mu_- + m_-\mu_+$, $\bar\lambda=m_+\lambda_- + m_-\lambda_+$ and
\begin{gather} \label{optimal translators domains}
\begin{split}
&\Psi_1^q=\left\{
							\bq_0: \begin{array}{l} \delta_2^q\leq2\delta_1^q\dfrac{\bar\mu+\mu_-+\bar\lambda}{2\mu_-+\bar\lambda}-1,
							\delta_2^q\leq 1,\\
                            \delta_1^q\leq2\delta_2^q\dfrac{\bar\mu+\mu_-+ \bar\lambda}{2\mu_-+\bar\lambda}-1,
                            \delta_1^q\leq 1 \end{array}
														\right\}, \\
&\Psi_2^q=\left\{
								\bq_0: \begin{array}{l} \delta_2^q\geq 2\delta_1^q\dfrac{\bar\mu+\mu_-+\bar\lambda}{2\mu_-+\bar\lambda}-1,
								\delta_2^q\geq\dfrac{\bar\lambda+2\mu_-}{\bar\lambda+2\bar\mu},\\
								\delta_2^q\leq 1,
								\delta_2^q\leq2\delta_1^q\dfrac{\bar\mu+\mu_-+\bar\lambda}{\bar\lambda}-1 \end{array}
								\right\},\\
&\Psi_3^q=\left\{\bq_0: \delta_1^q\geq\dfrac{\bar\lambda}{\bar\lambda+2\bar\mu},
										\delta_2^q\geq\dfrac{\bar\lambda}{\bar\lambda+2\bar\mu},
										\delta_1^q\leq\dfrac{\bar\lambda+2\mu_-}{\bar\lambda+2\bar\mu},
										\delta_2^q\leq\dfrac{\bar\lambda+2\mu_-}{\bar\lambda+2\bar\mu}
					\right\},
\\
&\Psi_4^q=\left\{\bq_0: \begin{array}{l} \delta_2^q\leq 1,
                             \delta_2^q\geq \dfrac{\bar\lambda(\bar\mu-\mu_-)\delta_1^q+ \bar\lambda(\bar\mu+\mu_-)+2\bar\mu\mu_-}{2\bar\mu(\bar\lambda+\bar\mu)},\\
                             \delta_2^q\geq2\delta_1^q\dfrac{\bar\mu+ \mu_-+\bar\lambda}{\bar\lambda}-1,
                             \delta_1^q\geq -1 \end{array}
                             \right\},\\
&\Psi_5^q=\left\{\bq_0: \begin{array}{l}\delta_2^q\leq \dfrac{\bar\lambda(\bar\mu-\mu_-)\delta_1^q+ \bar\lambda(\bar\mu+\mu_-)+2\bar\mu\mu_-}{2\bar\mu(\bar\lambda+\bar\mu)},\\
                             \delta_2^q\geq \dfrac{\bar\lambda(\delta_1^q+1)}{2(\bar\lambda+\bar\mu)},
                             \delta_1^q\leq\dfrac{\bar\lambda}{\bar\lambda+2\bar\mu},
                             \delta_1^q\geq -1 \end{array}
                             \right\}, \\
&\Psi_6^q=\left\{\bq_0: \begin{array}{l} \delta_2^q\leq \dfrac{\bar\lambda(\delta_1^q+1)}{2(\bar\lambda+\bar\mu)},
                             \delta_1^q\leq -1,\\
                             \delta_2^q\geq \dfrac{\delta_1^q(\bar\lambda(\mu_-+ \bar\mu)+2\mu_-\bar\mu)+\bar\lambda(\bar\mu-\mu_-)}{2\bar\mu(\bar\lambda+\bar\mu)}\end{array}
                             \right\},\\
&\Psi_7^q=\left\{\bq_0: \begin{array}{l} \delta_2^q\geq \delta_1^q,
                             \delta_1^q\geq -1, \\
                             \delta_2^q\leq \dfrac{\delta_1^q(\bar\lambda(\mu_-+ \bar\mu)+2\mu_-\bar\mu)+\bar\lambda(\bar\mu-\mu_-)}{2\bar\mu(\bar\lambda+\bar\mu)}   \\
							\end{array}
                             \right\}.
\end{split}
\end{gather}

\begin{figure}[!h]
\begin{center}
\SetLabels
\L (0.81*0.45)  { $\delta_1^{q}$} \\
\L (0.425*0.97) { $\delta_2^{q}$} \\
\L (0.52*0.67) { $\Psi_3^{q}$} \\
\L (0.66*0.87) { $\Psi_2^{q}$} \\
\L (0.73*0.78) { $\Psi_2^{q}{'}$} \\ %
\L (0.8*0.93) { $\Psi_1^{q}$} \\
\L (0.33*0.83) { $\Psi_4^{q}$} \\
\L (0.72*0.36) { $\Psi_4^{q}{'}$} \\
\L (0.18*0.6) { $\Psi_5^{q}$} \\
\L (0.57*0.35) { $\Psi_5^{q}{'}$} \\
\L (0.4*0.1) { $\Psi_6^{q}{'}$} \\
\L (0.18*0.32) { $\Psi_6^q$} \\
\L (0.3*0.3) { $\Psi_7^q$} \\
\L (0.76*0.9) \rule{6mm}{0.4mm} \\
\L (0.5*0.77) { A} \\
\L (0.51*0.93) { B} \\
\L (0.73*0.93) { C} \\
\L (0.66*0.77) { D} \\
\L (0.51*0.505) { E} \\
\L (0.125*0.73) { F} \\
\L (0.125*0.46) { G} \\
\L (0.125*0.23) { H} \\
\endSetLabels
\AffixLabels{\centerline{\includegraphics[width=0.8\textwidth]{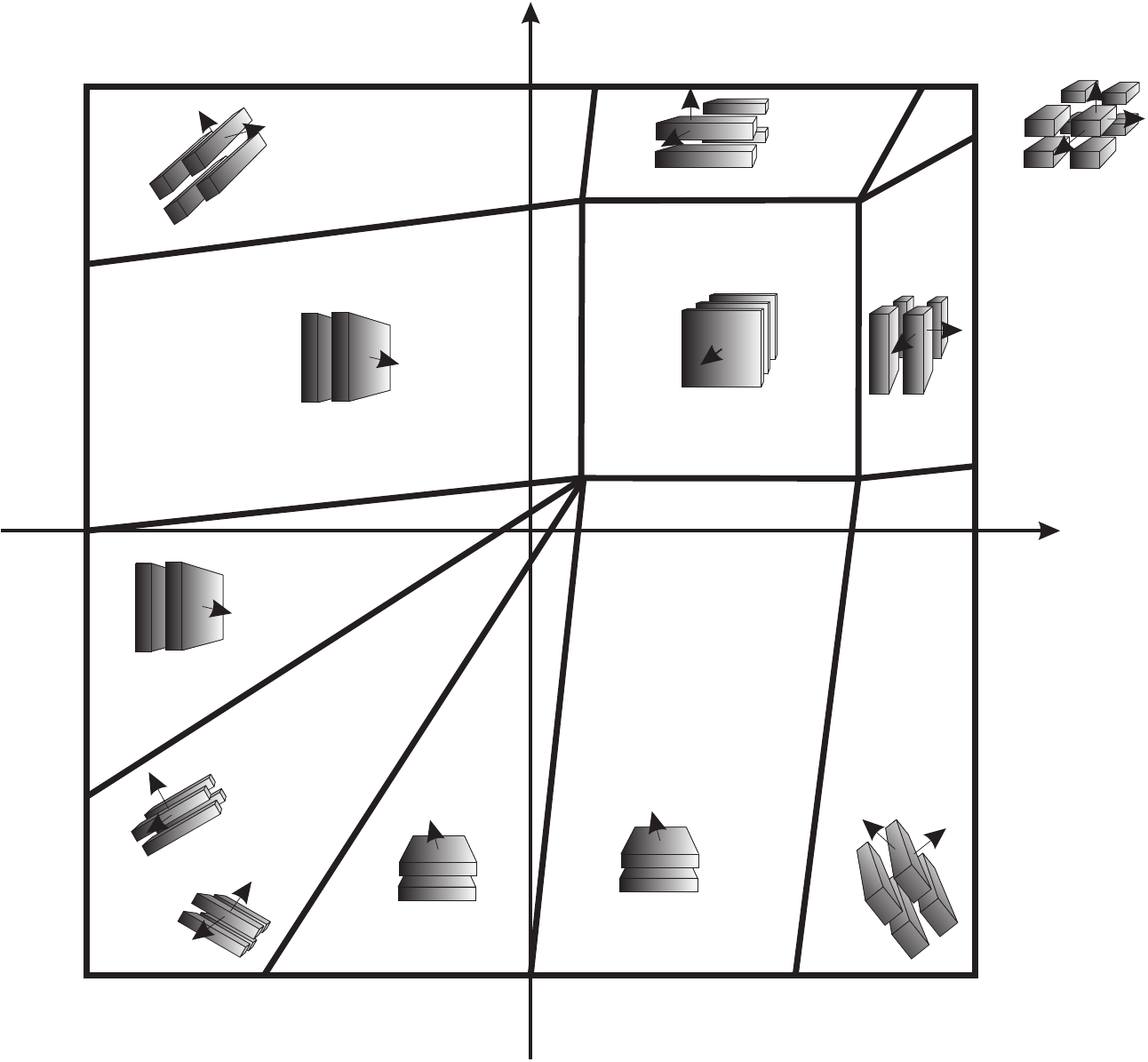}}}
\caption{\label{kvadrat-q} Optimal microstructures in dependence on  $\delta_1^q$ and $\delta_2^q$}
\end{center}
\end{figure}
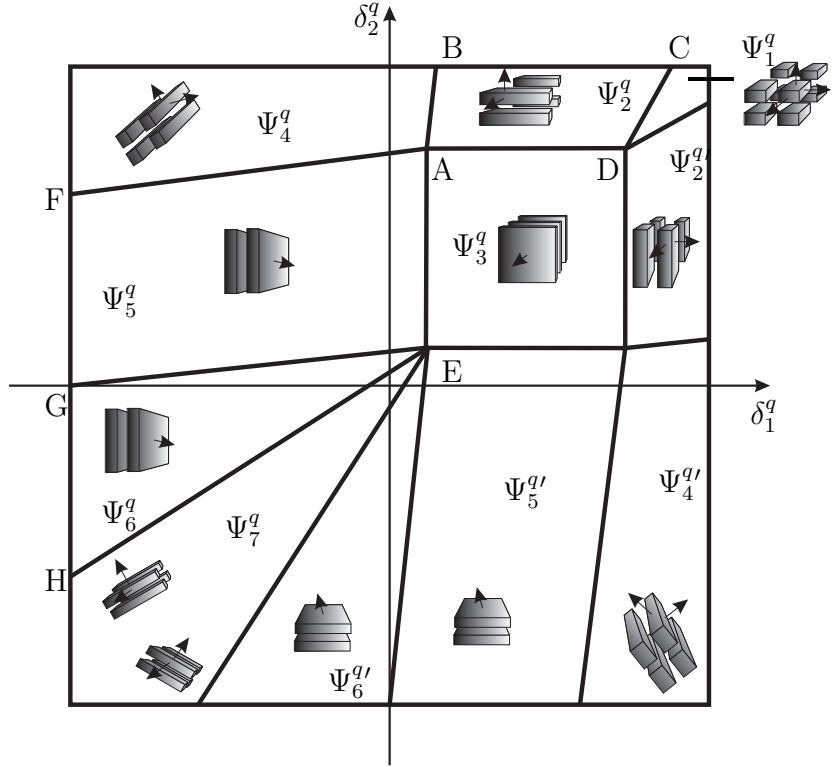

 In the domain $\Psi_1^q$   the tensor $\bq_+$  is spherical and all eigenvalues of the strain difference tensor $\Delta\obeps$ have the same sign. The  microstructures are third-rank laminates.

 In the domain $\Psi_2^q$  the tensor  $\bq_+$  is axisymmetric, ${\bq_+=q_k\be_1\be_1+q(\bE-\be_1\be_1)}$, the differences of the eigenvalues of $\obeps_\pm$ are such that $\Delta\overline\varepsilon_1=0$ and ${\Delta\overline\varepsilon_2\Delta\overline\varepsilon_3>0}$. The  microstructures are direct second-rank laminates.

 In the domains $\Psi_4^q$ and $\Psi_7^q$  the tensor $\bq_+$ is axisymmetric and the eigenvalues of the  tensor $\Delta\obeps$ differ from zero and have different signs. Thus, a vector $\btau$ exists such that $\Delta\obeps:\btau\btau=0$. The   microstructures are skew second-rank laminates.
In the domain $\Psi_4^q$  \ $\bq_+=q_k\be_1\be_1+q(\bE-\be_1\be_1)$, one of the normals lies in the plane passing  through the eigenvectors $\be_1$ and $\be_2$ of $\beps_0$   and the other normal lies in the plane passing through the eigenvectors $\be_1$ and $\be_3$.
In the domain $\Psi_7^q$   $\bq_+=q_k\be_3\be_3+q(\bE-\be_3\be_3)$,   one normal  lies in the plane passing through the vectors  $\be_1$ and $\be_3$ and the other one lies in the plane passing through $\be_2$ and $\be_3$.

 In the domain $\Psi_3^q$,  tensor $\bq_+$ has different eigenvalues and the eigenvalues of the  tensor $\Delta\overline{\varepsilon}$ are  $\Delta\overline{\varepsilon}_1=\Delta\overline{\varepsilon}_2=0$, $\Delta\overline{\varepsilon}_3\neq 0$. The  microstructures  are  simple laminates with the normal $\bn=\be_3$.

In the domains $\Psi_5^q$ and $\Psi_6^q$ the eigenvalues of the strain difference tensor are   $\Delta\overline{\varepsilon}_2=0$ and $\Delta\overline{\varepsilon}_1\Delta\overline{\varepsilon}_3<0$. The   microstructures are  simple laminates  with the normals lying in the plane passing through $\be_1$ and $\be_3$.

Other cases are obtained by permutation of indices. The domains and microstructures are shown in Fig.~\ref{kvadrat-q}.

\subsection{The attainability of the lower bound}

We prove that the lower bound
\begin{equation*}
	\mathcal{PW}=m_-w^t_-(\bt,\obeps_-)+m_+w^t_+(\bt,\obeps_+)+\varphi(\bt,\beps_0),
\end{equation*}
where $\bt$ is determined by \eqref{optimal translators} in domains \eqref{optimal translators domains}, can be attained by the optimal laminates, i.e. Eq. \eqref{exactness-of-bonds}
\begin{equation}\nonumber
	\mathcal{PW}=\mathcal{J}_{wm}=\mathcal{LW}
\end{equation}
holds if the strains averaged over the phases in the laminate equal to translation-optimal strains  $\obeps_-$ and $\obeps_+$.

At first, we note that since the jumps of strains across the interfaces in optimal laminates are defined by the relationships like \eqref{Kq-dif} or \eqref{Kq-dif_dirrect}, and the translator parameters are defined in \eqref{optimal translators}, then
the equality holds
\begin{equation} \label{lem}
\int\limits_{\Omega} \varphi(\bt, \beps)d\Omega = \varphi(\bt,\beps_0).
\end{equation}

Secondly, from the transmitting formula  \eqref{eq_eps_dif} it follows that the local strains in the phase ``$-$'' on every sublevel are related with the strain $\obeps_-$ as
\begin{equation} \nonumber
\beps_1^{(j)}=\obeps_-+\sum\limits_{i=2}^{n}\alpha_i^{(j)}\left(\bK_-(\bn^{(i-1)})-\bK_-(\bn^{(i)})\right):\bq_+,
\end{equation}
where parameters $\alpha_i^{(j)}$ depend on the volume fractions  of the sublayers on all levels (we do not need their explicit expressions).

 In the domains $\Psi_1^q, \Psi_2^q, \Psi_4^q$ and $\Psi_7^q$ where the second- and third-rank laminates are optimal, the translator is determined by tensors $\bt=2\mu_-\bE$, ${\bt=(2\mu_--t)\be_1\be_1+t\bE}$, ${\bt=2\mu_-\be_1\be_1}$ and ${\bt=2\mu_-\be_3\be_3}$, respectively. Therefore the  Hessian \eqref{Hessian} has zero eigenvalues in these domains, and
 \begin{equation}\label{sec_4_eq_104}
	(\bC_--\bT):\left(\bK_-(\bn^{(i-1)})-\bK_-(\bn^{(i)})\right):\bq_+=0
\end{equation}
in the case of optimal laminates.
Then from \eqref{wtT} and \eqref{sec_4_eq_104} it follows that the difference $\beps_1^{(j)}-\obeps_-$ is such that \begin{equation} \label{average-def}
	w_-^t(\bt,\beps_1^{(j)})=w_-^t(\bt,\obeps_-)
\end{equation}
where the fact is also used that the transformation strain is spherical.

Now, assuming that the average strains in the laminates are $\obeps_-$ and $\obeps_+$, we show that   the energy of the optimal laminates
\begin{equation}	\label{LW4}
\mathcal{LW}=\sum\limits_{i=1}^{n}\dfrac{m_1^{(i)}}{m_2^{(i)}}\prod\limits_{j=i}^{n}m_2^{(j)}w_-(\beps_1^{(i)})+m_+w_+(\obeps_+)
\end{equation}
equals to the lower bound.

We add and substract $\varphi(\bt,\beps_0)$ to the right-hand side of \eqref{LW4} and, using \eqref{lem}, bring $\mathcal{LW}$ to the form
\begin{equation*}	\mathcal{LW}= \sum\limits_{i=1}^{n}\dfrac{m_1^{(i)}}{m_2^{(i)}} \prod\limits_{j=i}^{n}m_2^{(j)}w_-^t(\bt,\beps_1^{(i)})+m_+w_+^t(\bt,\obeps_+)+\varphi(\bt,\beps_0).
\end{equation*}
The we express the strains $\beps_1^{(i)}$ through the average $\obeps_-$ and, using the relationship \eqref{average-def} and the equality ${m_-=\sum\limits_{i=1}^{n}\dfrac{m_1^{(i)}}{m_2^{(i)}}\prod\limits_{j=i}^{n}m_2^{(j)}}$,  obtain
\begin{equation}\nonumber
	\mathcal{LW}
=m_-w_-^t(\bt,\obeps_-)+m_+w_+^t(\bt,\obeps_+)+\varphi(\bt,\beps_0),
\end{equation}
which is equal to $\mathcal{PW}$.

\subsection{Optimal energy}
The common values of the upper $\mathcal{LW}(m_+, \beps_0)$ and lower $\mathcal{PW}(m_+, \beps_0)$ bounds of the energy define the optimal energy $\mathcal{J}_{wm}(m_+, \beps_0)$ as it is noted in \eqref{exactness-of-bonds}. The formulas for the optimal energy $\mathcal{J}_{wm}(m_+, \beps_0)$ are obtained by substitution of optimal $t_i$ into expression for $\mathcal{PW}$. The resulting formulas, however, are bulky and not instructive. We show them here  for a special representative case of zero eigenstrain $\beps_\pm^p=0$ and zero Poisson's ratio:

\begin{equation*}
	\begin{tabular}{ll}
		$\mathcal{J}_{wm}=\dfrac{\mu_-(\mu_++2\mu_-)}{\bar\mu+2\mu_-}(\varepsilon_1+\varepsilon_2+\varepsilon_3)^2-2\mu_-(\varepsilon_1\varepsilon_2+\varepsilon_2\varepsilon_3+\varepsilon_1\varepsilon_3)$ &\text{ if $\bq_0\in\Psi_1^q$}\\[4mm]	
		$\mathcal{J}_{wm}=(\mu_++\mu_--\bar\mu)\varepsilon_1^2+\dfrac{\mu_-(\mu_++\mu_-)}{\bar\mu+\mu_-}(\varepsilon_2+\varepsilon_3)^2-2\mu_-\varepsilon_2\varepsilon_3$ &\text{ if $\bq_0\in\Psi_2^q$}\\[4mm]	
		$\mathcal{J}_{wm}=(\mu_++\mu_--\bar\mu)(\varepsilon_1^2+\varepsilon_2^2)+\dfrac{\mu_-\mu_+}{\bar\mu}\varepsilon_3^2$ &\text{ if $\bq_0\in\Psi_3^q$}\\[4mm]	
		$\mathcal{J}_{wm}=\dfrac{\mu_-\mu_+}{\bar\mu}\varepsilon_1^2+\dfrac{\mu_-(\mu_++\mu_-)}{\bar\mu+\mu_-}(\varepsilon_2+\varepsilon_3)^2-2\mu_-\varepsilon_2\varepsilon_3$ &\text{ if $\bq_0\in\Psi_4^q$}\\[4mm]	
		$\mathcal{J}_{wm}=\dfrac{\mu_-\mu_+}{\bar\mu}(\varepsilon_1^2+\varepsilon_3^2)+(\mu_++\mu_--\bar\mu)\varepsilon_2^2$ &\text{ if $\bq_0\in\Psi_5^q\cup\Psi_6^q$}\\[4mm]	
		$\mathcal{J}_{wm}=\dfrac{\mu_-(\mu_++\mu_-)}{\bar\mu+\mu_-}(\varepsilon_1+\varepsilon_2)^2+\dfrac{\mu_-\mu_+}{\bar\mu}\varepsilon_3^2-2\mu_-\varepsilon_1\varepsilon_2$ &\text{ if $\bq_0\in\Psi_7^q$}\\
	\end{tabular}
\end{equation*}
where the domains $\Psi_i^q$ are derived easily from   \eqref{optimal translators domains} (see also \cite{AntimonovCherkaevFreidin2010,AntimonovNTV-2010Bounds}).

\subsection{Comments on optimal regimes}

1. The boundaries of the domains $\Psi^q_i$, derived from the optimality of the translated energy, coincide with the lines of  change of the type of optimal microstructures. Namely, at the boundary $CD$ the structure parameter $\beta^{(1)}$ in the third-rank laminates tends to zero and the third-rank laminates degenerate into second-rank laminates. At the boundaries $AF$, $AD$ and $EH$ one of the structure parameters of the second-rank laminates tends to zero and the second-rank laminates degenerate into simple laminates. At the boundaries $AB$ and $AE$, when moving from $\Psi_1^q$ to $\Psi_6^q$ or from $\Psi_2^q$ to $\Psi_4^q$, the condition \eqref{normal-1} holds and correspondingly, direct laminates are turned into skew laminates.

2. In the degenerative case $m_+=0$ the arrangement of the domains $\Psi_i^q$ does not change in general, but in the opposite case $m_-=0$ the boundaries $AF$ and $AD$ move to the boundary $\delta_2^q=1$ and the point $D$ tends to the corner point $\delta_1^q=\delta_2^q=1$. This implies that only simple laminates are optimal when the concentration of the softer phase $m_-$  tends to zero.

3. We have shown that minimal energy at any external strain corresponds to an optimal laminate structure. The minimal energy of the optimal microstructures forms a multi-faced surface that is explicitly described by simple analytic formulas. Thus, we obtained a parametric representation for the quasiconvex envelope of two-well Lagrangian for elastic energy in 3D and have described the minimizing sequences (fields in optimal laminates) for this problem.

\section{Phase transformations limit surfaces}

The phase transformations limit surfaces are formed by the strains at which phase transformation can occur at first time on a given strain path. This means that the microstructure with an infinitesimal volume fraction of one of the phases has a minimal energy in comparison with other microstructures.

Recall that the energy of an elastic body undergoing phase transformations is determined by \eqref{energy min form3} as
\begin{equation}\nonumber
\mathcal{J}_f(\beps_0)=\inf\limits_{\begin{minipage}{17mm}\scriptsize\centerline{$0 \leq\! m_+ \leq 1$} \end{minipage}} \mathcal{F}_m
\end{equation}
where the minimum of the free energy at given external strains and volume fractions of the phases is
\begin{equation} \label{optimal_free_energy}
\mathcal{F}_{m}=f_{-}^0+m_+\gamma +\mathcal{J}_{wm}(m_+,\beps_0).
\end{equation}
and $\mathcal{J}_{wm}(m_+,\beps_0)$ is the optimal strain energy that is exact lower bound of the strain energy  at given $m_+$   and $\beps_0$.

Considering phase transformation requires an additional minimization of \eqref{optimal_free_energy} with respect to the volume fraction $m_+$. Since the optimal strain energy coincides with the energy of one of the optimal laminates, it is convenient to minimize the energy using the lamination formulae. As a result the dependencies $m_+(\beps_0)$ and stress-strain diagrams can be constructed on the path of the phase transformation at $0 \leq m_+ \leq 1 $ (see, e.g., example of such a strategy for the case of simple laminates in \cite{FreidinSharipova2006}).

Further we focus on the transformation limit surfaces construction
 for the direct and reverse phase transformations.  After taking the interaction energy
$$
E=m_+(\gamma_*+\dfrac12 \be_0:\bM_-^{(n)}:\be_0), \  n=1,2,3
$$
 of the optimal laminates (i.e. of the optimal rank $n$ and normals) instead of $\mathcal{F}_{m}$, we obtain that the limit surfaces are formed by all  strains $\beps_0$ at which one of the minimizing volume fractions tends to zero,
\begin{gather}
\label{reverse}
\left.\dfrac{\partial E(\beps_0,m_+)}{\partial m_+}\right|_{m_+=0}=0,
\end{gather}
or to unit,
%
\begin{gather}\label{ur_pr_pov}
\left.\dfrac{\partial E(\beps_0,m_+)}{\partial m_+}\right|_{m_+=1}=0.
\end{gather}

The initial undeformed state corresponds to the phase with a lower chemical energy $f^0$. We will refer to this phase as to a parent phase. Note also that we accept that $\mu_-<\mu_+$. 
Therefore,
if the shear module of the parent phase is greater than the shear module of a new phase,
then the parent and new phases are  phases ``$+$'' and ``$-$'', respectively; $\gamma=f_+^0-f_-^0<0$. Eq. \eqref{ur_pr_pov} determines the direct transformation limit surface (from ``$+$'' to ``$-$''), and Eq. \eqref{reverse} determines the reverse transformation limit surface (from ``$-$'' to ``$+$'').

If the shear module of the parent phase is less than the shear module of the new phase,
 then the parent and new phases are phases ``$-$'' and ``$+$'', respectively; $\gamma=f_+^0-f_-^0>0$. The direct transformation surface for the transformation from ``$-$'' to ``$+$'' is determined by Eq. \eqref{reverse}, and Eq. \eqref{ur_pr_pov} determines the reverse transformation limit surface for the transformation from ``$+$'' to ``$-$''.

 It can be derived (see Appendix~E) that for all optimal laminates
\begin{equation}\label{Maxw1}
\dfrac{\partial E}{\partial m_+}=\gamma_*+\dfrac12 (\bq_+:\sk\bC^{-1}:\bq_++\mathcal{K}_{max}(\bq_+))=0
\end{equation}
where the dependence of $\mathcal{K}_{max}$ on the eigenvalues of $\bq_+$ is given by one of the formulae \eqref{Kq-dif1}, \eqref{Kmax}, \eqref{qaxi2-3}, \eqref{Kmax2} or \eqref{cK}.

Eq. \eqref{Maxw1} is a restriction on $\bq_+$ at all equilibrium volume fractions~$m_+$. This is the Maxwell relation  for the equilibrium interface expressed in terms of strains on one side of the interface \cite{Kublanov1988} that additionally corresponds  to the external PTZ boundary, i.e. with the strain $\beps_+$ belonging to the PTZ boundary (see, e.g., \cite{MorozovFreidin1998,FreidinVilchSharTAM2002,FreidinSharipova2006} or \cite{Freidin2007} and references therein).

The substitution of $\bq_+=\bM^{(n)}:\sk{\bC}^{-1}:\bq_0$ with the optimal rank, normals and $m_+=0$ or $m_+=1$ into \eqref{Maxw1} results in the equation of the limit surfaces in the space of eigenvalues $q_{0i}$ of the tensor $\bq_0=\sum\limits_{i=1}^3q_{0i}\be_i\be_i$.

The choice of the rank and the normals depends on external strains.  The space of eigenvalues of $\bq_0$ is   divided into domains which correspond to domains  $\Psi_i^q$ ($i=1,2...7$). Different optimal laminates and different relationships between the tensor $\bq_+$ and external strains $\beps_0$ (different tensors $\bM_-^{(n)}$) correspond to different domains $\Psi_i^q$. Thus, the transformation  surfaces are combinations of parts which correspond to the nucleation of different optimal laminates.

If $m_+\rightarrow 1$ then $m_-\rightarrow 0$ and, by \eqref{qplusM} and \eqref{Mn}, $\bq_+=\bq_0$.  Eq. \eqref{Maxw1} becomes the equation of the limit surface of the nucleation of a phase ``$-$'', the phase with the less shear module. The space is divided into domains corresponding $\Psi_3^q$, $\Psi_5^q$ and $\Psi_6^q$ since only first-rank laminates are optimal, $\mathcal{K}_{max}$ is determined by \eqref{Kq-dif} or \eqref{Kmax} (see also comment 2 in Subsection 4.7).
The limit surface  is formed by second-order surfaces in strain space. Note that the limit surface in this case coincides with the external PTZ boundary.

Other optimal laminates may correspond to the nucleation of the phase ``$+$'' ($m_+\rightarrow 0$), the phase with a greater shear module.
For example, in the case of the optimal skew second-rank laminates $\bq_+=(q_k-q)\bk\bk+q\bE$. Eq.~\eqref{Maxw1} takes the form
\begin{equation}\label{Maxw2}
	2\gamma_*+\dfrac{(q_k+2q)^2}{9\sk k}+\dfrac{(q_k-q)^2}{3\sk \mu} + \mathcal{K}_{max}=0
\end{equation}
where $\mathcal{K}_{max}$ is given by \eqref{qaxi2-3}.

By \eqref{qplusM}, the relation between $\bq_0$ and $\bq_+$ is
$$
\bq_0=\{\bI+m_- \sk\bC :(\beta \bK_-(\bn^{(1)})+(1-\beta)\bK_-(\bn^{(2)}))\}:\bq_+$$
where $\beta=m_1^{(1)}/m_-$. For the optimal skew second-rank laminates $\bk$ is one of the eigenvectors of~$\bq_0$.
Let $\bk=\be_1$. Then from \eqref{qaxi2} and \eqref{qaxi2-1} it follows at $m_+=0$  ($m_-=1$) that
\begin{gather} \label{sk1}
	q_{01}=q_k+\dfrac{1}{2\mu_-}\left(\sk\lambda (q_k-q)+2\sk\mu((1-\nu_-)q_k-\nu_-q)\right), \\ \label{sk2}
	q_{02} = q+\dfrac{1}{2\mu_-}\left(\sk\lambda (q_k-q)+2\beta\sk\mu((1-\nu_-)q-\nu_-q_k)\right),\\ \label{sk3}
	q_{03} = q+\dfrac{1}{2\mu_-}\left(\sk\lambda (q_k-q)+2(1-\beta)\sk\mu((1-\nu_-)q-\nu_-q_k)\right).
\end{gather}

From \eqref{sk2} and  \eqref{sk3} it follows that
\begin{equation} \label{sk4}
	q_{02}+q_{03} = 2q+\dfrac{1}{2\mu_-}\left(2\sk\lambda (q_k-q)+2\sk\mu((1-\nu_-)q-\nu_-q_k)\right).
\end{equation}

Given $q_{01}$, $q_{02}$ and $q_{03}$, equations \eqref{sk1} and \eqref{sk4} determine $q$ and $q_k$ in dependence on $q_{01}$ and $q_{02}+q_{03}$. After substitution of these dependencies into \eqref{Maxw2} we obtain the equation of a cylindrical surface in space of $q_{01}$, $q_{02}$ and $q_{03}$. Thus, the skew second-rank laminate nucleation surface is a part of the cylindrical surface restricted by the domain $\Psi_4^q$. Boundaries of $\Psi_4^q$ correspond to $\beta=0$ and $1$ and  to the strains when $q_k=0$.

Other optimal laminates can be considered analogously. The parts of the limit surface that correspond to the first-rank laminates are parts of the ellipsoids or hyperboloids
\begin{equation}\nonumber
\gamma_*+\dfrac12 (\bq_0:{\sk\bC}^{-1}:\bq_0-\mathcal{K}^+_{max}(\bq_0))=0
\end{equation}
where $\mathcal{K}^+_{max}(\bq_0)$ is the maximum of $\bq_0:\bK_+(\bn):\bq_0$, where $\bK_+$ is defined by the elasticity tensor $\bC_+$ \cite{FreidinSharipova2006}. The limit surfaces for the direct second-rank laminates are cylinders.
The limit surfaces for the third-rank laminates are the triangles, as for the ellipsoidal nuclei \cite{Kublanov1988,Freidin2007,Freidin2009} (see Appendix~E for details).

\section{Results and discussion}

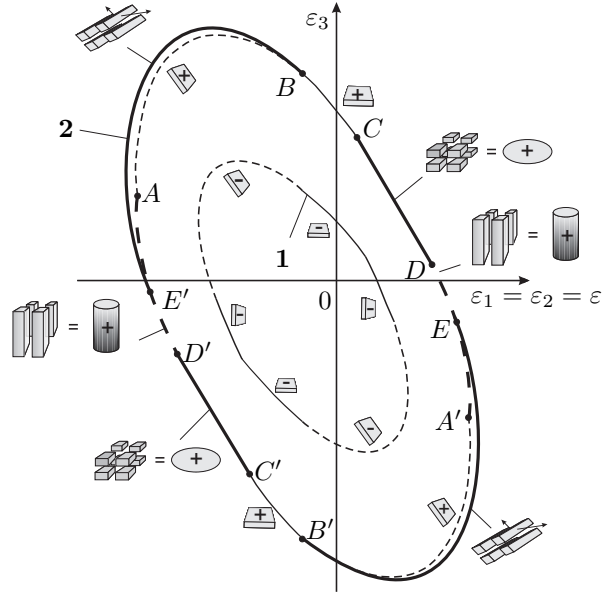
\begin{figure}[tb]
\begin{center}\footnotesize
\SetLabels
\L (0.51*0.96) {$\varepsilon_3$} \\
\L (0.67*0.49) {$\varepsilon_1=\varepsilon_2=\varepsilon$} \\
\L (0.35*0.661) $A$ \\
\L (0.635*0.275) $A'$ \\
\L (0.48*0.55) {$\mathbf{1}$} \\
\L (0.27*0.77) {$\mathbf{2}$} \\
\L (0.48*0.84) $B$ \\
\L (0.565*0.77) $C$ \\
\L (0.605*0.53) $D$ \\
\L (0.63*0.43) $E$ \\
\L (0.37*0.48) $E'$ \\
\L (0.39*0.39) $D'$ \\
\L (0.46*0.19) $C'$ \\
\L (0.51*0.09) $B'$ \\
\L (0.522*0.48) $0$ \\
\endSetLabels
\begin{center}
\AffixLabels{\centerline{\includegraphics[width=0.55\textwidth]{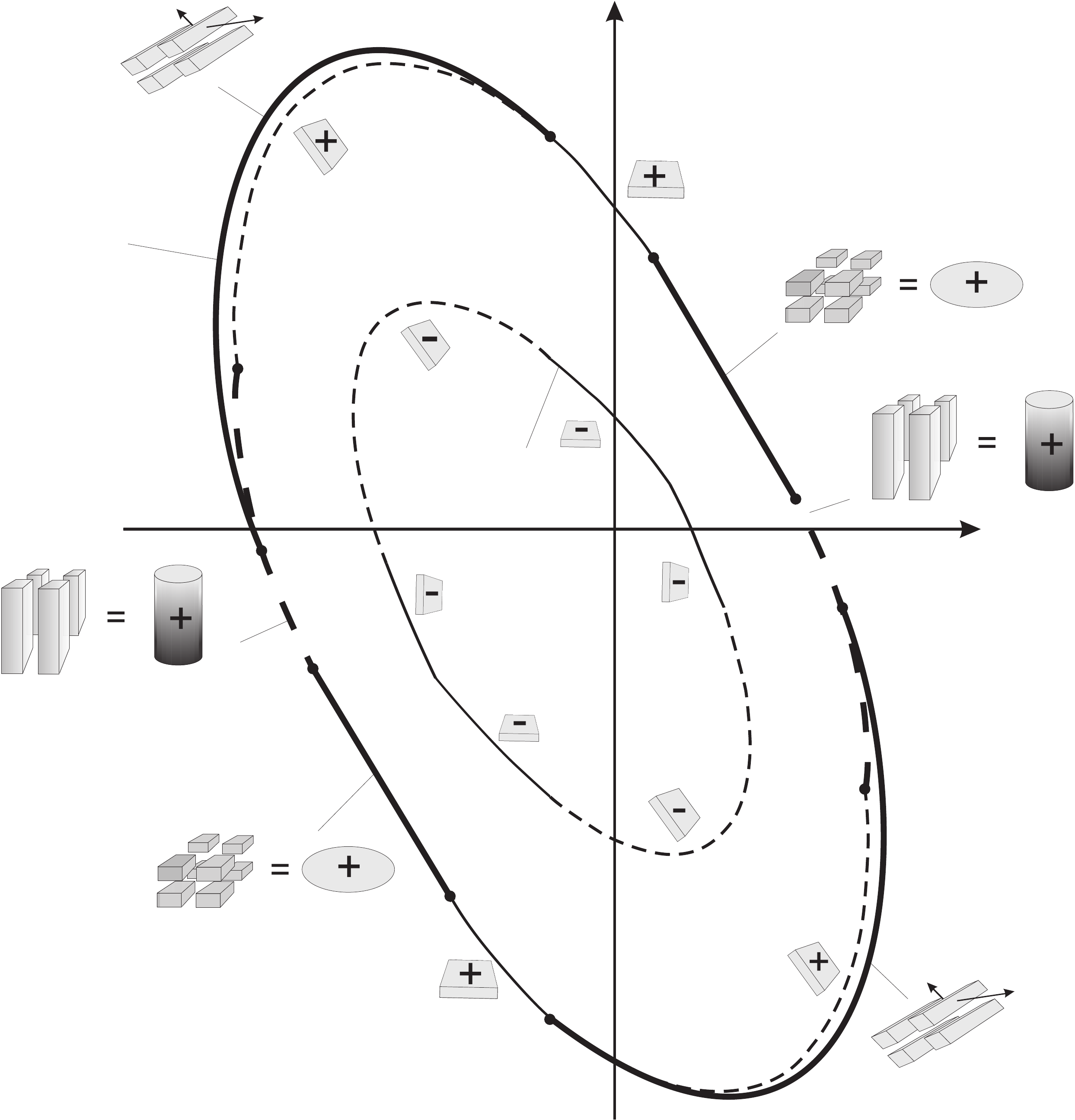}}}
\end{center}
\end{center}
\caption{Cross sections of phase transformation surfaces by plane $\varepsilon_1=\varepsilon_2$. Material parameters are $f^0_--f^0_+=0.5,\,\ k_+=78,\,\
k_-=39,\,\ \mu_+=30,\,\ \mu_-=15,\,\
\eps^p_+=0, \ \varepsilon^p_-=0.1$
\label{Transform1}}
\end{figure}

\begin{figure}[!tb]
\begin{center}\footnotesize
\SetLabels
\L (0.34*0.98) {$\varepsilon_3$} \\
\L (0.7*0.24) {$\varepsilon_1=\varepsilon_2=\varepsilon$} \\
\L (0.35*0.67) $A$ \\
\L (0.635*0.29) $A'$ \\
\L (0.48*0.56) {$\mathbf{2}$} \\
\L (0.27*0.79) {$\mathbf{1}$} \\
\L (0.485*0.855) $B$ \\
\L (0.53*0.76) $C$ \\
\L (0.604*0.54) $D$ \\
\L (0.626*0.44) $E$ \\
\L (0.33*0.47) $E'$ \\
\L (0.39*0.4) $D'$ \\
\L (0.46*0.18) $C'$ \\
\L (0.5*0.09) $B'$ \\
\L (0.358*0.233) $0$ \\
\endSetLabels
\begin{center}
\AffixLabels{\centerline{\includegraphics[width=0.55\textwidth]{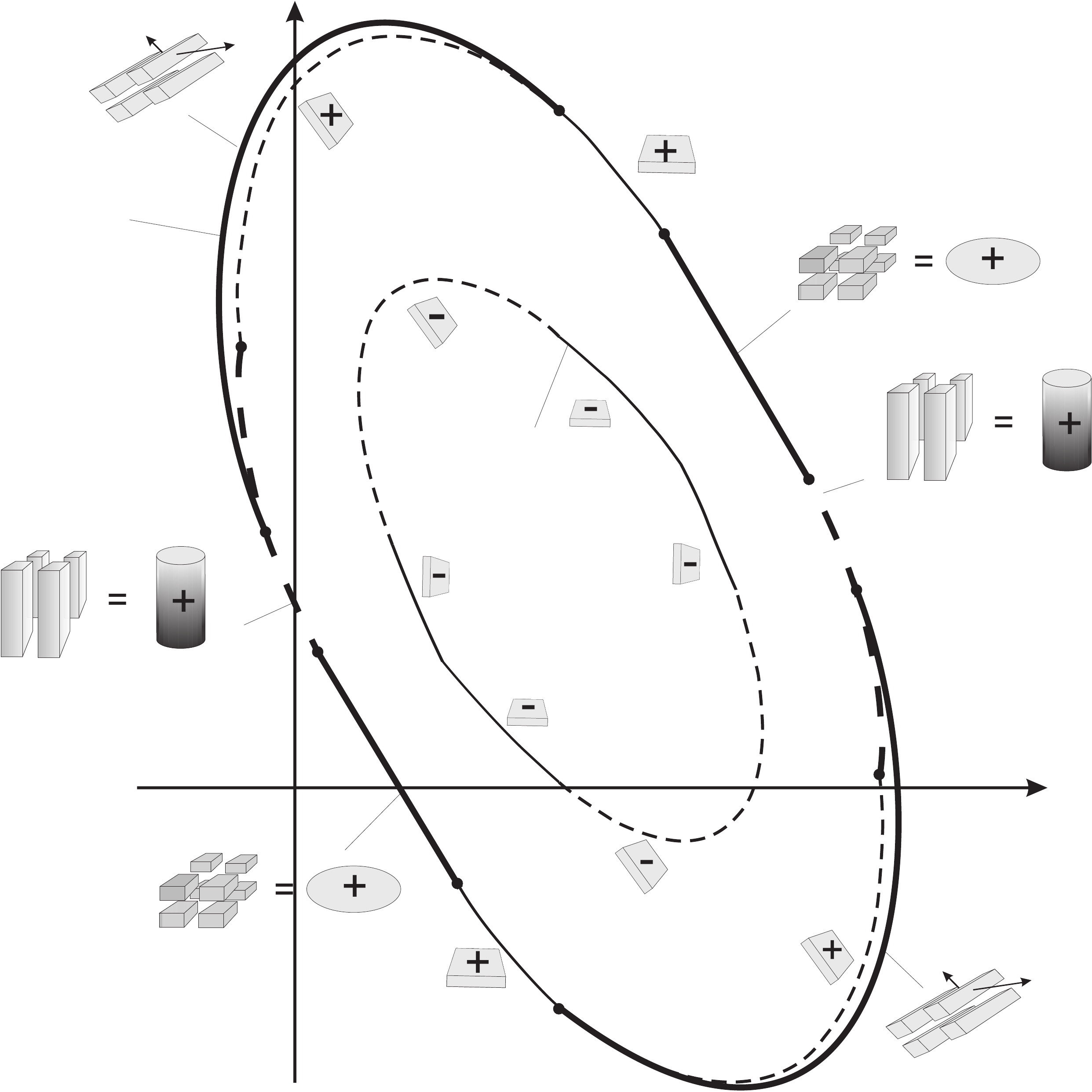}}}
\end{center}
\end{center}
\caption{Cross sections of phase transformation surfaces by plane $\varepsilon_1=\varepsilon_2$. Material parameters are $f^0_+-f^0_-=0.5,\,\ k_+=78,\,\
k_-=39,\,\ \mu_+=30,\,\ \mu_-=15,\,\
\eps^p_+=0.1, \ \varepsilon^p_-=0$
\label{Transform2}}
\end{figure}

\begin{figure}[!tb]
\begin{center}\footnotesize
\SetLabels
\L (0.453*0.97) {$\varepsilon_3$} \\
\L (0.64*0.505) {$\varepsilon_1=\varepsilon_2=\varepsilon$} \\
\L (0.335*0.645) $A$ \\
\L (0.585*0.33) $A'$ \\
\L (0.63*0.41) {$\mathbf{1}$} \\
\L (0.295*0.193) {$\mathbf{2}$} \\
\L (0.414*0.753) $C'$ \\
\L (0.442*0.64) $B'$ \\
\L (0.51*0.21) $C$ \\
\L (0.445*0.29) $B$ \\
\L (0.465*0.503) $0$ \\
\endSetLabels
\begin{center}
\AffixLabels{\centerline{\includegraphics[width=0.5\textwidth]{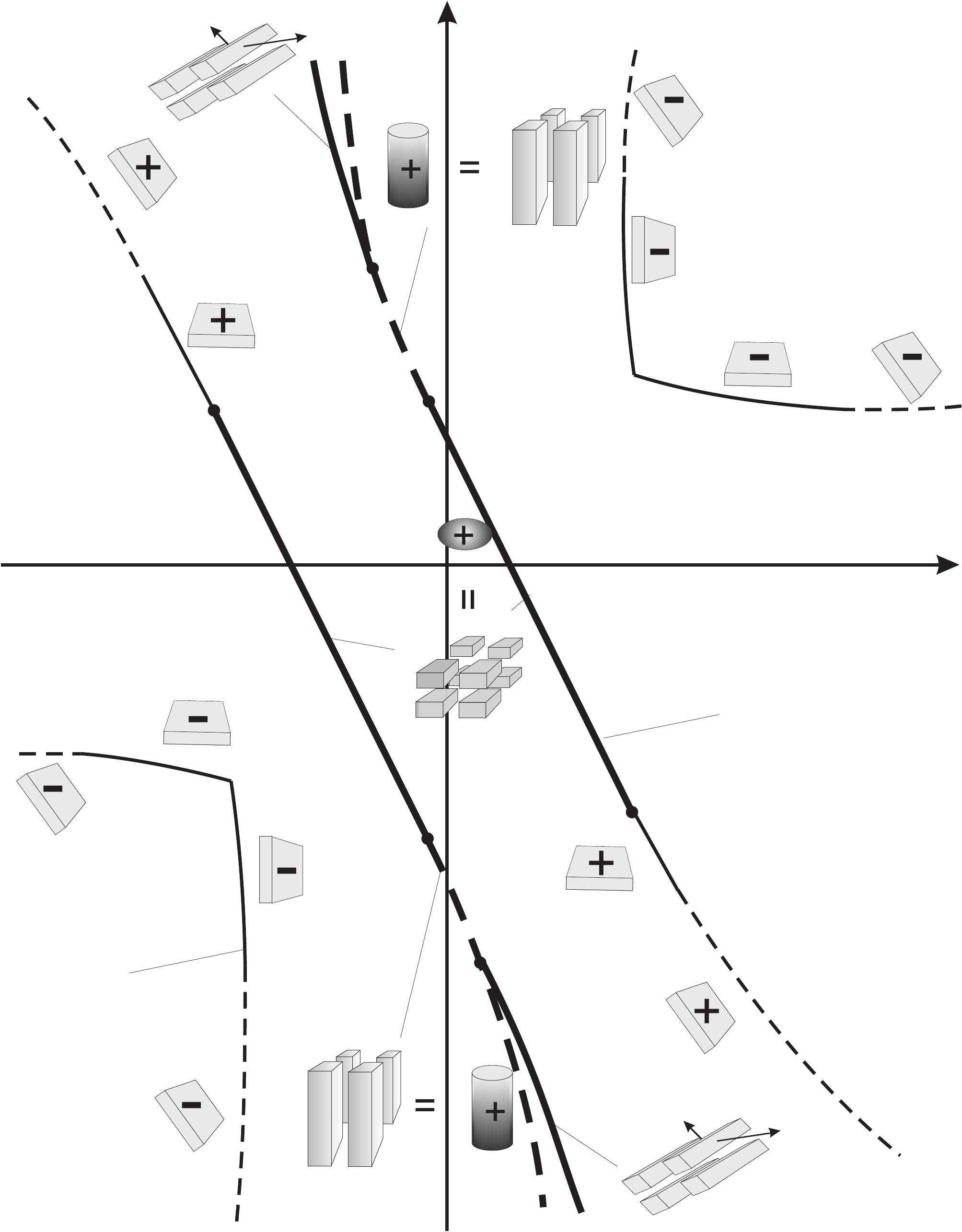}}}
\end{center}
\end{center}
\caption{Cross sections of phase transformation surfaces by plane $\varepsilon_1=\varepsilon_2$. Material parameters are $f^0_+-f^0_-=0.5,\,\ k_+=39,\,\
k_-=78,\,\ \mu_+=30,\,\ \mu_-=15,\,\
\eps^p_+=0.1, \ \varepsilon^p_-=0$
\label{Transform3}}
\end{figure}

\begin{figure}[!tb]
\begin{center}\footnotesize
\SetLabels
\L (0.28*0.97) {$\varepsilon_3$} \\
\L (0.64*0.245) {$\varepsilon_1=\varepsilon_2=\varepsilon$} \\
\L (0.335*0.65) $A$ \\
\L (0.545*0.32) $A'$ \\
\L (0.63*0.415) {$\mathbf{2}$} \\
\L (0.67*0.54) {$\mathbf{1}$} \\
\L (0.415*0.75) $C'$ \\
\L (0.48*0.67) $B'$ \\
\L (0.505*0.22) $C$ \\
\L (0.45*0.29) $B$ \\
\L (0.297*0.24) $0$ \\
\endSetLabels
\begin{center}
\AffixLabels{\centerline{\includegraphics[width=0.5\textwidth]{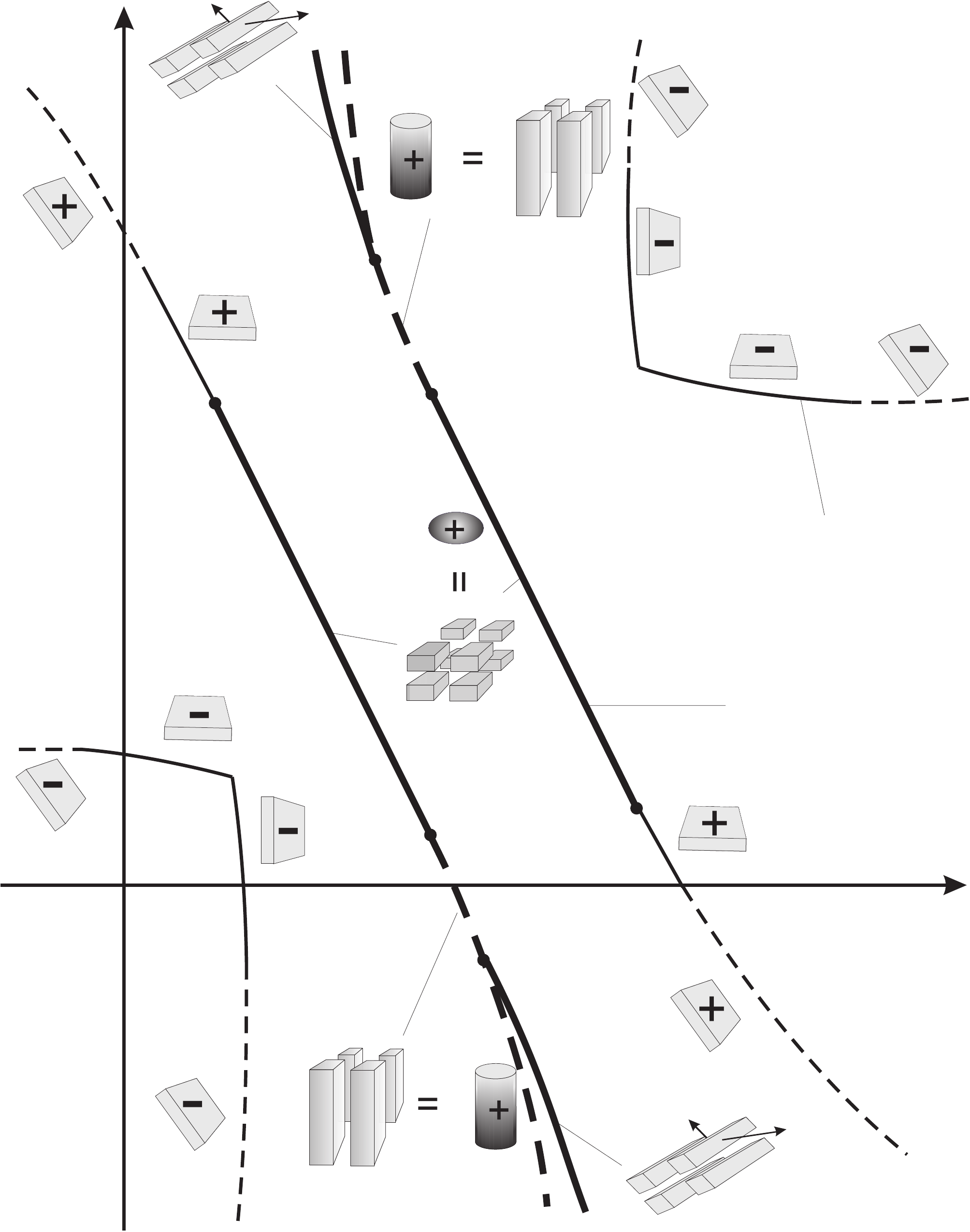}}}
\end{center}
\end{center}
\caption{Cross sections of phase transformation surfaces by plane $\varepsilon_1=\varepsilon_2$. Material parameters are $f^0_--f^0_+=0.5,\,\ k_+=39,\,\
k_-=78,\,\ \mu_+=30,\,\ \mu_-=15,\,\
\eps^p_+=0, \ \varepsilon^p_-=0.1$
\label{Transform4}}
\end{figure}

Transformation surfaces are formed by second-order surfaces (cylindrical, ellipsoidal or hyperboloidal, depending on the sign-definiteness of the jump $\sk{\bC}$) and planes (for the third-rank laminates).
Examples of the cross-sections of the direct and reverse transformation limit surfaces by the plane $\varepsilon_1=\varepsilon_2$ are shown in Figs.~\ref{Transform1}--\ref{Transform4}. The cross-sections correspond  to axisymmetric external strains. The material parameters are indicated in the captions.
The curves~``1'' are the direct transformation surfaces, the curves~``2'' are the reverse transformation surfaces.

Also, envelope of the nucleation surfaces for equilibrium nuclei of different geometries (planar layers, elliptical cylinders, ellipsoids) with interfaces which satisfy the local thermodynamic equilibrium condition (the Maxwell relation) obtained earlier \cite{Kublanov1988,FreidinVilchSharTAM2002,FreidinSharipova2006,Freidin2009,
Antimonov2009,AntimonovNTV-2010Cylinder} are shown.

Given transformation strain and the ``temperature'' $\gamma$, the position of the transformation surfaces with respect to the origin of coordinates (zero external strain) and the shape of the limit surfaces (closed or unclosed) depend on the relation between elastic moduli of the phases. This in turn affects the type of the microstructures that appear on the path of direct and reverse transformations.

The surfaces are closed if the jump of the elasticity tensor is sign-definite (Figs.~\ref{Transform1}~and~\ref{Transform2}) and unclosed if the jump is not a sign-definite tensor
(Figs.~\ref{Transform3} and \ref{Transform4}).

If the shear module of the parent phase is greater than the shear module of a new phase then the limit surfaces surround the coordinate origin (Fig.~\ref{Transform1}). The transformation takes place at both stretching and compression. When the direct transformation surface (curve ``1'' in Fig.~\ref{Transform1}) is reached on the straining path from the unstrained state, the layers of a new phase ``$-$'' nucleate inside the parent phase ``$+$''. The orientation of the layer depends on the strain state, according to formulae \eqref{normal-1}--\eqref{qaxi4}. It may be a ``direct'' laminate with layers oriented perpendicular to the direction of maximal stretch like a normal crack, or an inclined laminate like shear bands, and the angle depends on the strains.

The reverse transformation $\mbox{``$-$''}\rightarrow \mbox{``$+$''}$ occurs during unloading.
The reverse transformation limit surface (the curve 2) corresponds to third-rank laminates equivalent to the ellipsoidal nuclei, simple direct laminates, direct second-rank laminates equivalent to cylinder nuclei and skew second-rank laminates. The skew second-rank laminates replace direct second-rank laminates (points $E$ and $E'$) or simple laminates (points $B$ and $B'$). This means that the equilibrium cylinders or layers do not provide the minimum of energy in this part of strain space, even if they satisfy thermodynamic equilibrium conditions and their local instability was not found. We emphasize that that skew second-rank laminates make the transformation surface convex.

If the shear module of the parent phase is less than the shear module of a new phase then the limit surfaces are shifted with respect to the coordinates origin (Fig.~\ref{Transform2}). As opposite to the previous case, there are linear straining paths which start from the origin and do not cross the direct transformation limit surface; no direct phase transformation in such cases whatever the values of strains are. The free energy of one-phase state is always less than the energy of two-phase state at these paths.
Third-rank laminates (equivalent to the ellipsoids), direct (equivalent to cylinders) and skew second-rank laminates and simple laminates (simple layers) correspond to the direct transformation limit surface in this case.
Direct and inclined simple laminates correspond to the reverse transformation limit surface.

Note that strains inside the phase ``$+$'' are spherical in the third-rank laminates as well as inside equilibrium ellipsoid. Therefore the energy is not affected by increasing the shear module in the case of ellipsoids nucleation. This explains different geometry of microstructures on the paths of direct and reverse transformations at different relations of the shear moduli.

If the jump of the elasticity tensor is not sign-definite then the transformation surface are unclosed, opposite to the case of the sign-definite jump of the elasticity tensor (Figs.~\ref{Transform3} and \ref{Transform4}). As it is in the case of the sign-definite jump of the elasticity tensor, one of the limit surfaces corresponds to simple laminates, another one to the laminates of the first, second and third ranks, and one can see that the skew second-rank laminates affect the transformation limit surfaces.

\section{Conclusions}

In the present paper we constructed phase transformations limit surfaces in strain space basing on exact energy lower bounds for elastic two-phase composites and demonstrated that the optimal laminates construction may be an effective tool for transformation surfaces construction. These surfaces define external strains at which a two-phase microstructure minimizes the free energy at an infinitesimal volume fraction of one of the phases. The case of isotropic phases was considered. To construct the limit surfaces we solved two problems. At first, the sequential laminates with minimal energy were studied in detail. Then the lower bound was constructed basing on translation method. The coincidence of the energy of optimal laminates and the lower bound was demonstrated. We also noted that the minimizing strains in one of the phases must belong to the external PTZ-boundary that is in agreement with earlier observations (Eremeev et al., 2007; Freidin et al., 2006; Fu \& Freidin, 2004; Grabovsky \& Truskinovsky, 2011, 2013).

The solution showed that there are new regimes of optimality of microstructures in comparison with the asymptotic cases considered in other papers, namely, so-called skew second-rank laminates are the minimizers in some part of strain space. The case of the optimal skew laminates also complements solutions  considered in Chenchiah and Bhattacharya (2008).

It was also shown that the transformation limit surfaces only partly coincide with the envelope of the nucleation surfaces constructed earlier by the semi-inverse method for planar, cylindrical and ellipsoidal nuclei. It was noted that such an envelope may be nonconvex and namely skew second-rank laminates, which do not correspond to any of above-mentioned nuclei, make the transformation surface convex in such a case.

We showed that the limit surfaces can be closed or unclosed, depending on the 
 sign-definiteness of the jump of the elasticity tensor. We also demonstrated that microstructures of various types correspond to the direct and reverse transformations, and the type of the microstructure depends on the relation between shear moduli of the phases.

\section*{Acknowledgments}
Andrej Cherkaev is thankful to NSF for the support through the grant of NSF (grant DMS-1515125). Alexander Freidin greatly appreciates the  support of Russian Foundation for Basic Research (grant no. 13-01-00687).

\appendix
\section{Transmitting tensor $\bM_-^{(n)}$}
Below we derive the expression \eqref{Mn} for the transmitting tensor $\bM^{(n)}$ and prove the Proposition~\ref{prop1} about its  sign-definiteness.

\emph{A1. Derivation of the transmitting tensor $\bM_-^{(n)}$}

We start with considering \emph{rank-1 laminates} which
consist   of simple layers 1 and 2. Let $m_1^{(1)}=m_-$ and $m_2^{(1)}=m_+$ are the volume fractions of the layers occupied by the materials ``$-$'' and ``$+$'', $\bn^{(1)}$ is the normal to the layers, $\beps_1^{(1)}$, $\beps_2^{(1)}$ and $\bsig_1^{(1)} $,
$\bsig_2^{(1)}$ are the strains and stresses inside the layers $1$ and $2$. The following relationships are satisfied:
\begin{gather}
	\label{eq_1}
m_1^{(1)}\beps_1^{(1)}+m_2^{(1)}\beps_2^{(1)}=\beps_2^{(2)}, \quad
	m_1^{(1)}\bsig_1^{(1)}+m_2^{(1)}\bsig_2^{(1)}=\bsig_2^{(2)},
\\
\label{eq_2}
\bsig_1^{(1)}=\bC_- :(\beps_1^{(1)}-\beps_-^p), \quad  \bsig_2^{(1)}=\bC_2^{(1)}: (\beps_2^{(1)}- \beps_2^{p(1)} ),\\
\label{eq_lam1}
\bC_2^{(1)}=\bC_+, \quad \beps_2^{p(1)} =\beps^p_+.
\end{gather}
Average strain $\beps_0=\beps_2^{(2)}$ and stress,  $\bsig_0=\bsig_2^{(2)}$ are related as
\begin{equation}\label{00}
\bsig_0=\bC_0:(\beps_0-\beps^p_0)
\end{equation}
 where  effective elasticity tensor
$\bC_0=\bC_2^{(2)}$ and effective transformation strain $\beps^p_0=\beps_2^{p(2)}$  are to be found.
Note that in the case of higher-rank laminates $\beps_2^{(2)}$ and $\bsig_2^{(2)}$ are unknown  strain and stress averaged within  the rank-1 sublayer. That is why we keep super- and subscripts denoting the rank and number of a layer even in the case of the rank-1 laminate.

Following relationships for the rank-1 laminate were presented in \cite{FreidinSharipova2006}. We repeat the derivations to have the notation consistent with further considerations of higher ranks laminates.
From \eqref{eq_1} and \eqref{eq_02} it follows that
\begin{equation}\label{eq_4}
\beps_2^{(2)}=\beps_2^{(1)}-m_1^{(1)}(\beps_2^{(1)}-\beps_1^{(1)}) =\beps_2^{(1)}+m_1^{(1)}\bK_-(\bn^{(1)}):\bq_2^{(1)}
\end{equation}
where
\begin{equation*}
\bq_2^{(1)} =  \left(\bC_2^{(1)}-\bC_1^{(1)}\right):\beps_2^{(1)} -\left(\bC_2^{(1)}:\beps_2^{p (1)}\!-\bC_1^{(1)}:\beps_1^{p(1)}\right).
\end{equation*}

Equation \eqref{eq_4} can be rewritten as
\begin{equation}\nonumber
\left(\left(\bC_2^{(1)}-\bC_1^{(1)}\right)^{-1}+m_1^{(1)}\bK_-(\bn^{(1)})\right):\bq_2^{(1)}=\be_2^{(2)}
\end{equation}
where
\begin{equation*}
\be_2^{(2)}= \beps_2^{(2)}-\left(\bC_2^{(1)}-\bC_1^{(1)}\right)^{-1} :\left(\bC_2^{(1)}:
\beps_2^{p(1)}-\bC_1^{(1)}:\beps_1^{p(1)}\right).
\end{equation*}
Then
\begin{equation}\label{eq_5}
	\bq_2^{(1)}=\bL_-^{(1)}(\bn^{(1)},m_1^{(1)}):\be_2^{(2)}
\end{equation}
where
\begin{equation*}
\bL_-^{(1)}(\bn^{(1)},m_1^{(1)})= \left((\bC_2^{(1)}-\bC_1^{(1)})^{-1}+m_1^{(1)}\bK_-(\bn^{(1)})\right)^{-1}.
\end{equation*}

By \eqref{eq_lam1},
\begin{gather}\nonumber
\bq_2^{(1)} =\sk{\bC}:\beps_2^{(1)}-\sk{\bC:\beps^p}= \bq_+,\\
\label{e22-0}
\be_2^{(2)}= \beps_2^{(2)}-\sk{\bC}^{-1}:\sk{\bC:\beps^p},\\
\nonumber 
\bL_-^{(1)}(\bn^{(1)},m_1^{(1)})= \left(\sk{\bC}^{-1}+m_1^{(1)}\bK_-(\bn^{(1)})\right)^{-1}.
\end{gather}
 Since $\beps_2^{(2)}=\beps_0$, from \eqref{e22-0} it follows that
\begin{equation}\nonumber
\be_2^{(2)}=\beps_0-\sk{\bC}^{-1}:\sk{\bC:\beps^p}\equiv \be_0.
\end{equation}
Finally \eqref{eq_5} takes the form of the dependence of $\bq_+$
on external strain  and microstructure parameters $m_1^{(1)}$, $\bn^{(1)}$:
\begin{equation}\label{q+eps01}
\bq_+=\bL_-^{(1)}(\bn^{(1)},m_1^{(1)}):\be_0.
\end{equation}
To relate $\bsig_2^{(2)}$ and $\beps_2^{(2)}$,  we note that from \eqref{eq_1} and \eqref{eq_2} it follows that
\begin{equation}\label{eq_7}
\bsig_2^{(2)}=\bC_-:(\beps_2^{(2)}-\beps_-^p)+m_2^{(1)}\bq_2^{(1)}.
\end{equation}
After substituting \eqref{q+eps01} into \eqref{eq_7} we derive that
\begin{equation*}
\bsig_2^{(2)}=\bC_2^{(2)}:\left(\beps_2^{(2)}- \beps_2^{p(2)}\right)
\end{equation*}
where the effective elasticity tensor and  effective transformation strain of the rank-1 laminate
\begin{gather}
\bC_2^{(2)}=\bC_-+m_2^{(1)}\bL_-^{(1)},
\label{eq_8}
\\
\label{eq_9}
	\left(\beps_2^p\right)^{(2)}=\beps_-^p+ m_2^{(1)}\left(\bC_2^{(2)}\right)^{-1}:\bL_-^{(1)}:\sk{\bC}^{-1}:\bC_+:\sk{\beps^p}.
\end{gather}
Further we will use the tensors \eqref{eq_8} and \eqref{eq_9} as the elasticity tensor and transformation tensor   of a homogenized rank-1 sublayer of higher rank laminates.

\emph{Rank-2 laminates} are characterized by the volume fraction $m_1^{(2)}$ of the simple  layers of the phase ``$-$'' and the volume fraction $m_2^{(2)}$ of the mixed layers which themselves are rank-1 laminates characterized by the volume fractions $m_1^{(1)}$ and $m_2^{(1)}$ of the sublayers ``$-$'' and ``$+$''; $\bn^{(2)}$ and $\bn^{(1)}$ are the normals to the macrolayers and to sublayers inside rank-1 layers, respectively.

The volume fractions satisfy  the relationships:
\begin{equation*}
	m_1^{(2)}+m_2^{(2)}=1, \quad m_1^{(1)}+m_2^{(1)}=1, \quad m_2^{(1)}m_2^{(2)}=m_+.
\end{equation*}
Then
\begin{equation*}
	m_1^{(2)}+m_1^{(1)}m_2^{(2)} = m_1^{(1)}+m_1^{(2)}m_2^{(1)}=m_-.
\end{equation*}

Strains $\beps_1^{(2)}$, $\beps_1^{(1)}$, $\beps_2^{(2)}$,   $\beps_2^{(1)}$ and stresses $\bsig_1^{(2)}$, $\bsig_1^{(1)}$, $\bsig_2^{(2)}$, $\bsig_2^{(1)}$ satisfy the relationships
\begin{gather}\nonumber
	m_1^{(1)}\beps_1^{(1)}+m_2^{(1)}\beps_2^{(1)}=\beps_2^{(2)}, \quad m_1^{(2)}\beps_1^{(2)}+m_2^{(2)}\beps_2^{(2)}=\beps_2^{(3)}=\beps_0,\\
\nonumber
m_1^{(1)}\bsig_1^{(1)}+m_2^{(1)}\bsig_2^{(1)}=\bsig_2^{(2)}, \quad m_1^{(2)}\bsig_1^{(2)}+m_2^{(2)}\bsig_2^{(2)}=\bsig_2^{(3)}=\bsig_0,\\
\nonumber
\bsig_1^{(1)}=\bC_- :(\beps_1^{(1)}-\beps_-^p), \quad  \bsig_2^{(1)}=\bC_2^{(1)}:(\beps_2^{(1)}-\beps_2^{p(1)}),\\
\nonumber
\bsig_1^{(2)}=\bC_- :(\beps_1^{(2)}-\beps_-^p), \quad  \bsig_2^{(1)}=\bC_2^{(2)}:
(\beps_2^{(2)}-\beps_2^{p(2)} ),
\end{gather}
where $\bC_2^{(1)}=\bC_+$, \; $\beps_2^{p(1)} =\beps^p_+$, \; $\bC_2^{(2)}$ and $\beps_2^{p(2)}$ are defined by \eqref{eq_8} and~\eqref{eq_9}.

Eq. \eqref{eq_5} expresses $\bq_2^{(1)}=\bq_+$ through $\be_2^{(2)}$ defined by \eqref{e22-0}. Now we have to express $\be_2^{(2)}$ through $\beps_0$.
Considering rank-$1$ sublayer as a homogenized layer with the effective elasticity tensor
$\bC_2^{(2)}$ and effective transformation strain  $ \beps_2^{p(2)}$ defined by $\eqref{eq_8}$ and $\eqref{eq_9}$ we derive similar to \eqref{eq_5} that
\begin{equation}\label{eq_13a}
	\bq_2^{(2)}=\bL_-^{(2)}\left(\bn^{(2)},m_1^{(2)}\right):\be_2^{(3)},
\end{equation}
where by the definition
\begin{gather}\label{eq_13}
\bq_2^{(2)}= \left(\bC_2^{(2)}-\bC_-\right):\beps_2^{(2)} -\left(\bC_2^{(2)}:\beps_2^{p(2)}-\bC_- :\beps_-^p\right),\\
\label{eq_17}
	\be_2^{(3)}= \beps_0-\left(\bC_2^{(2)}-\bC_-\right)^{-1}:\left(\bC_2^{(2)}: \beps_2^{p(2)}-\bC_-:\beps_-^p\right),\\
\label{Lminus2}
\bL_-^{(2)}= \left(\left(\bC_2^{(2)}-\bC_-\right)^{-1}+m_1^{(2)}\bK_-(\bn^{(2)})\right)^{-1}.
\end{gather}

From \eqref{eq_8} and \eqref{eq_9} it follows that
\begin{gather}\nonumber
\bC_2^{(2)} -\bC_- =m_2^{(1)}\bL_-^{(1)},\\
\nonumber
	\bC_2^{(2)}:\beps_2^{p(2)}-\bC_-:\beps_-^p= \left(\bC_2^{(2)}-\bC_-\right):\beps_-^p+\bC_2^{(2)}: \left(\beps_2^{p (2)}-\beps_-^p\right) \quad\qquad \\ \nonumber 
 \quad =m_2^{(1)}\bL_-^{(1)}:\left(\beps_-^p+\sk{\bC}^{-1}:\bC_+:\sk{\beps^p}\right) =m_2^{(1)}\bL_-^{(1)}:\sk{\bC}^{-1}:\sk{\bC:\beps^p},\\
 \nonumber
(\bC_2^{(2)} -\bC_-)^{-1}:(\bC_2^{(2)}:\beps_2^{p(2)}-\bC_-:\beps_-^p)=\sk{\bC}^{-1}:\sk{\bC:\beps^p}
\end{gather}
Then, by \eqref{eq_13}--\eqref{Lminus2}, with $\be_2^{(2)}$ defined by \eqref{e22-0} and with the use of \eqref{eq_5},
\begin{gather}\label{eq_17}
\bq_2^{(2)}= m_2^{(1)}\bL_-^{(1)}:\be_2^{(2)}=m_2^{(1)}\bq_2^{(1)}=m_2^{(1)}\bq_+,\\
\nonumber
\be_2^{(3)}=\beps_0-\sk{\bC}^{-1}:\sk{\bC:\beps^p}=\be_0,\\
\nonumber
\bL_-^{(2)}={m_2^{(1)}} \left(\sk{\bC}^{-1}+m_1^{(1)}\bK_-(\bn^{(1)})+m_1^{(2)}m_2^{(1)}\bK_-(\bn^{(2)})\right)^{-1}.
\end{gather}
From \eqref{eq_17} and \eqref{eq_13a} it follows that
\begin{gather*}
\bq_+ = 
\dfrac{1}{m_2^{(1)}}\bq_2^{(2)} =\bM_-^{(2)}:\be_0 
\end{gather*}
where
\begin{equation*}
	\bM_-^{(2)}= \dfrac1{m_2^{(1)}}\bL_-^{(2)}= \left(\sk{\bC}^{-1}+m_1^{(1)}\bK_-(\bn^{(1)})+m_1^{(2)}m_2^{(1)}\bK_-(\bn^{(2)})\right)^{-1}.
\end{equation*}

Average stress $\bsig_0=\bsig_2^{(3)}$ and strain $\beps_0=\beps_2^{(3)}$ are related by \eqref{00}
with the effective elasticity tensor and effective transformation strain
\begin{gather*}
\bC_0=\bC_2^{(3)}=\bC_-+m_2^{(2)}\bL_-^{(2)}=\bC_-+m_+\bM_-^{(2)},\\
	\beps_2^{p (3)}=\beps_-^p+
m_+\left(\bC_2^{(3)}\right)^{-1}:\bM_-^{(2)}:\sk{\bC}^{-1}:\bC_+:\sk{\beps^p}.
\end{gather*}

Considering rank-3 laminates, we derive that $\bq_2^{(3)}=m_2^{(2)}\bq_2^{(2)}=m_2^{(2)}m_2^{(1)}\bq_+$
and finally, for rank-n laminates, come to the formula \eqref{q2i-qplus} and then to the formulae   \eqref{qplusM}--\eqref{beta} and \eqref{effective}. Note that ${\bf M}_-^{(1)}=\bL_-^{(1)}$ and \eqref{qplusM}  reduces to \eqref{q+eps01} at $n=1$.

\emph{A2. The proof of the Proposition~\ref{prop1} (the sign of $\bM^{(n)}$)}

Obviously, if $\sk{\bC}>0$ then $\bM^{(n)}>0$.
Let $\sk{\bC}<0$.   From the identities \cite{Kunin2}
\begin{equation*}
\bK_+:\sk{\bC}:\bK_-=\bK_-:\sk{\bC}:\bK_+=-\sk{\bK}
\end{equation*}
it follows that \cite{NazyrovFreidin}
\begin{equation*}
(\bI+\bK_-:\sk{\bC}):(\bI-\bK_+:\sk{\bC})=\bI.
\end{equation*}
This provides the existence of the inverse tensors $(\sk{\bC}^{-1}+\bK_-)^{-1}$ and $(\sk{\bC}^{-1}-\bK_+)^{-1}$. Particularly,
\begin{gather*}
(\sk{\bC}^{-1}+\bK_-)^{-1}=\sk{\bC}-\sk{\bC}:\bK_+:\sk{\bC}\leq\sk{\bC}<0
\end{gather*}
and, thus,
\begin{equation}\label{CK0}
\sk{\bC}^{-1}+\bK_-<0.
\end{equation}
Since $\beta^{(i)}<1$ and $m_-<1$, from \eqref{CK0} it follows that
\begin{equation}\label{CK1}
\beta^{(i)}\sk{\bC}^{-1}+m_-\beta^{(i)}\bK_-(\bn^{(i)})<0.
\end{equation}
Summation of the inequalities \eqref{CK1} from  $i=1$ till $i=n$ with  taking into account \eqref{beta} gives
\begin{equation*}
\sk{\bC}^{-1}+m_- \sum\limits_{i=1}^{n}\beta^{(i)}\bK_-(\bn^{(i)})<0
\end{equation*}
where from it follows that $\bM_-^{(n)}$ exists and $\bM_-^{(n)}<0$. 

\section{Energy equivalence of laminates with permutated normals. Rank reduction}

\emph{B1. Derivation of the transmitting formula \eqref{eq_eps_dif}}

  The difference $\beps_1^{(i)}-\beps_1^{(i-1)}$ can be written down as
\begin{gather}\label{dif1}
\beps_1^{(i)}-\beps_1^{(i-1)}= (\beps_1^{(i)}-\beps_2^{(i)})-(\beps_1^{(i-1)}-\beps_2^{(i-1)})+(\beps_2^{(i)}-\beps_2^{(i-1)}) \end{gather}
Since $\beps_2^{(i)}=m_1^{(i-1)}\beps_1^{(i-1)}+m_2^{(i-1)}\beps_2^{(i-1)}$ and $1-m_2^{(i-1)}=m_1^{(i-1)}$, the last summand transforms to
\begin{gather*}
\beps_2^{(i)}-\beps_2^{(i-1)}= m_1^{(i-1)}(\beps_1^{(i-1)}-\beps_2^{(i-1)})
\end{gather*}
and the representation \eqref{dif1} takes the form
\begin{gather}\label{dif2}
\beps_1^{(i)}-\beps_1^{(i-1)}= (\beps_1^{(i)}-\beps_2^{(i)})-m_2^{(i-1)}(\beps_1^{(i-1)}-\beps_2^{(i-1)}) \end{gather}
Then the transmitting formula \eqref{eq_eps_dif} follows from \eqref{dif2} if it is taken into account that
\begin{gather*}
\beps_1^{(i)}-\beps_2^{(i)}=\bK_-(\bn^{(i)}):\bq_2^{(i)}, \quad \beps_1^{(i-1)}-\beps_2^{(i-1)}=\bK_-(\bn^{(i-1)}):\bq_2^{(i-1)}, \\
\bq_2^{(i)}=\bq_+\prod\limits_{j=1}^{i-1}m_2^{(j)}, \quad
\bq_2^{(i-1)}=\bq_+\prod\limits_{j=1}^{i-2}m_2^{(j)}
\end{gather*}

\emph{B2. The proof of the Proposition~\ref{permute}}

Let $\bq_+^A$ and $\bq_+^B$ are the tensors $\bq_+$ in the laminates $A$ and $B$ with permuted normals $\bn^{(k)}$ and $\bn^{(k+1)}$, respectively.  Then,
by \eqref{qplusM} and \eqref{Mn},
\begin{multline}\label{MqA}
\bigg(\sk{\bC}^{-1}+m_- \!\!\!\!\!\!\sum\limits_{i=1, \; i\neq k,k+1}^{n}\!\!\!\!\!\beta^{(i)}\bK_-(\bn^{(i)})
\\
+ m_-\left(\beta_A^{(k)}\bK_-(\bn_A^{(k)})+\beta_A^{(k+1)}\bK_-(\bn_A^{(k+1)})\right)\bigg):\bq_+^A=\be_0,
\end{multline}
\begin{multline}
\label{MqB}
\bigg(\sk{\bC}^{-1}+m_- \!\!\!\!\!\!
\sum\limits_{i=1, \; i\neq k,k+1}^{n}\!\!\!\!\!\beta^{(i)}\bK_-(\bn^{(i)})
\\
+ m_-\left(\beta_B^{(k)}\bK_-(\bn_B^{(k)})+\beta_B^{(k+1)}\bK_-(\bn_B^{(k+1)})\right)\bigg):\bq_+^B=\be_0.
\end{multline}

If  $k\geq 2$  then, by \eqref{beta-def},
\begin{equation*}
\beta_{A,B}^{(k)}=\dfrac{m_{1A,B}^{(k)}}{m_-}\prod\limits_{j=1}^{j=k-1}m_2^{(j)}, \quad \beta_{A,B}^{(k+1)}=\dfrac{m_{1A,B}^{(k+1)}m_{2A,B}^{(k)}}{m_-}\prod\limits_{j=1}^{j=k-1}m_2^{(j)}.
\end{equation*}
Then due to \eqref{mAB}
\begin{equation}
\nonumber
\beta_B^{(k)}=\beta_A^{(k+1)}\quad \mbox{and}\quad \beta_B^{(k+1)}=\beta_A^{(k)}
\end{equation}
and, thus,
\begin{equation*}
\beta_A^{(k)}\bK_-(\bn_A^{(k)})+\beta_A^{(k+1)}\bK_-(\bn_A^{(k+1)})= \beta_B^{(k)}\bK_-(\bn_B^{(k)})+\beta_A^{(k+1)}\bK_-(\bn_A^{(k+1)}).
\end{equation*}

If $i<k$ then the coefficients $\beta^{(i)}$ do not depend on the volume fractions on the \mbox{$k^{th}$} and \mbox{$(k+1)^{th}$} levels.
If $i>k+1$ then the coefficients $\beta_{A,B}^{(i)}$ depend on the volume fractions on the $k^{th}$ and $(k+1)^{th}$ levels through the products $m_{2A,B}^{(k)}m_{2A,B}^{(k+1)}$.
From \eqref{mAB} it follows that
\begin{equation}\label{mm}m_{2B}^{(k)}m_{2B}^{(k+1)}=m_{2B}^{(k)}(1-m_{1B}^{(k+1)})=m_{2B}^{(k)}-m_{1A}^{(k)}= 1-m_{1A}^{(k+1)}m_{2A}^{(k)}-m_{1A}^{(k)}=
m_{2A}^{(k)}m_{2A}^{(k+1)}.\end{equation}
Thus, the product $m_{2 }^{(k)}m_{2 }^{(k+1)}$ does not change and
\begin{equation*}
\beta_B^{(i)}=\beta_A^{(i)}, \quad i\neq k, k+1
\end{equation*}
From \eqref{mm} it also follows that relationships \eqref{mAB} do not contradict  the restriction $\prod\limits_{i=1}^{n}m_{2B}^{(i)}=m_+$.

If $k=1$ then $\beta_{A,B}^{(1)}=\dfrac{m_{1A,B}^{(1)}}{m_-}$ and $\beta_{A,B}^{(2)}=\dfrac{m_1^{(2)}m_{2A,B}^{(1)}}{m_-}$. By \eqref{mAB},
 $\beta_B^{(1)}=\beta_A^{(2)}\quad \mbox{and}\quad \beta_B^{(2)}=\beta_A^{(1)}$. By \eqref{mm},  the product $m_{2}^{(1)}m_{2}^{(2)}$ does not change after the permutation of the normals. Thus,
 $\beta_B^{(i)}=\beta_A^{(i)}, \ i>2$.

 Finally from \eqref{MqA}, \eqref{MqB} follows that $\bM^{(n)}_{B-}=\bM^{(n)}_{A-}$, \; ${\bq_+^B=\bq_+^A\equiv\bq_+}$, and, by \eqref{eq:rank-n energy}, ${W_B^{(n)}=W_A^{(n)}}$.

\emph{B3. The proof of the Proposition~\ref{prop2}  (rank reduction)}

At first we consider the case
\begin{equation}\label{case}\bK_-(\bn^{(1)}):\bq_+=\bK_-(\bn^{(2)}):\bq_+.
\end{equation}
By the transmitting
formula~\eqref{eq_eps_dif}
\begin{gather*}
\beps_1^{(2)}-\beps_1^{(1)}= m_2^{(1)} \left(\bK_-(\bn^{(2)})-\bK_-(\bn^{(1)})\right):\bq_+=0.
\end{gather*}
Then  $\beps_1^{(1)}=\beps_1^{(2)}\equiv \beps_{1\ast}^{(1)}$. Therefore the strain energy will not change if one replaces the rank-2 sublayer of the rank-$n$ laminate by the rank-1 laminate with the strains $\beps_{1\ast}^{(1)}$ and $\beps_{2\ast}^{(1)}=\beps_+$ and the volume fractions $m_{1\ast}^{(1)}=m_1^{(2)}+m_1^{(1)}m_2^{(2)}$ and $m_{2\ast}^{(1)}=m_2^{(1)}m_2^{(2)}$. This in turn means that the rank-$n$ laminate can be replaced by the rank-$(n-1)$ laminate without changing the energy.

If $\bK_-(\bn^{(i)}):\bq_+=\bK_-(\bn^{(j)}):\bq_+$ then permutating sequentially the normal $\bn^{(i)}$ with the normals $\bn^{(i-1)}$,  $\bn^{(i-2)}$, etc.  and permutating the normal $\bn^{(j)}$ with  the normals $\bn^{(j-1)}$,  $\bn^{(j-2)}$, etc. until the  normals $\bn^{(i)}$, $\bn^{(j)}$ will appear at the first two levels, and changing the volume fractions by the rule \eqref{mAB} we will finally reduce \eqref{KqCorol} to the  case \eqref{case}.

\emph{B4. Optimal laminates of ranks higher than 3}

We show  that the optimal laminates with the rank higher than three
are energy equivalent to the optimal laminates with the rank not more than 3.
This means that in 3D-case  maximum required rank is 3.

Indeed, even if $n=3$, then only one case remains to be unreduced to lower rank laminates, namely the case 3(b) with spherical $\bq_+$ in Appendix C. So, only spherical tensors may be considered as trial
$\bq_+$  if ${n> 3}$, and the condition $\bq_+:\bK_-(\bn_i);\bq_+=\mathrm{const}\ i=1,2...n$ is fulfilled in this case for any set of normals. But the energy of the laminate does not depend on the rank if the tensor $\bq_+$ is spherical. Indeed, if $\bq_+=q\bE$ then, by \eqref{qplusM}, \eqref{KqE} and \eqref{beta}
\begin{gather*}
\be_0=\Big(
\sk{\bC}^{-1}:\bE +m_-c q
\sum\limits_{i=1}^{n}\beta^{(i)}\bn^{(i)}\otimes \bn^{(i)}
\Big),
\\
\tr\,\be_0=\bE:\sk{\bC}^{-1}:\bE+m_-cq.
\end{gather*}
Thus, if at given $\beps_0$ the tensor $\bq_+$ is spherical then $q$ does not depend on the rank.
Then, by \eqref{Eq-int}, the energy
\begin{equation*}
	E_q=\dfrac12 m_+ q \,\tr\,\be_0
\end{equation*}
also does not depend on the rank.

 \section{The choice of laminates}

 If we restrict the consideration by the laminates of the first, second and third rank than the following cases can be considered depending on the relations between the eigenvalues of the tensor $\bq_+$.

\emph{1. Rank-1 laminates.} This case is denoted as the case $I$ in Section 3. The normal is uniquely determined by the eigenvalues of $\bq$ at all $\bq_+$ from the
 condition of the maximum of the quadratic form $\mathcal{K_-(\bn, \bq_+)}$.

\emph{2. Rank-2 laminates.}
The conditions \eqref{zeta} and  \eqref{qKq_var} are to be satisfied at two  normals $\bn^{(1)}$ and $\bn^{(2)}$.
From \eqref{Kq-dif1}, \eqref{qaxi2-3}, \eqref{qaxi4} and \eqref{cK} one can see  that the equality
$$
\mathcal{K}_-(\mathbf{n}^{(1)},\mathbf{q}_+)=\mathcal{K}_-(\mathbf{n}^{(2)},\mathbf{q}_+)
=\max\limits_{\bn:\; |\bn|=1}\mathcal{K}_-(\bn\,|\,\bq_+ ), \quad  \mathbf{n}^{(2)}\nparallel
\mathbf{n}^{(1)}
$$
is possible in four cases:
\begin{itemize}
\item[(a)] The eigenvalues of $\bq_+$ are different, one of the inequalities \eqref{normal-1} holds and both normals are determined by Eq. \eqref{normal}, but  differ by the sign of one of the component:
$$
n_{max}^{(2)}=-n_{max}^{(1)}, \  n_{min}^{(2)}=n_{min}^{(1)}, \  n_{mid}^{(1,2)}=0
$$
or
$$
n_{max}^{(2)}=n_{max}^{(1)}, \  n_{min}^{(2)}=-n_{min}^{(1)}, \  n_{mid}^{(1,2)}=0
$$

By \eqref{Kq-dif}, in both cases not only
\begin{equation}\label{qKqrank2}
\bq_+ :\bK_-(\bn^{(1)} ):\bq_+=\bq_+ :\bK_-(\bn^{(2)} ):\bq_+
 \end{equation}
 but also
\begin{equation}\nonumber
\bK_-(\bn^{(1)} ):\bq_+=\bK_-(\bn^{(2)} ):\bq_+
\end{equation}
Thus, by the Proposition~\ref{prop2}, the rank can be lowered without changing the strain energy, and the case 2(a) can be reduced to the rank-1 laminate.

 \item[(b)] The tensor $\bq_+$ is axisymmetric, $\bq_+=q(\bE-\bk\otimes\bk)+q_k\bk\otimes\bk$, and the eigenvalues $q$ and $q_k$ satisfy \eqref{qaxi1}.
     Two unit unequal normals have to satisfy the equality \eqref{qKqrank2}.
     Then any two different normals with equal projections onto the axis $\bk$ determined by \eqref{qaxi2}  can be taken as $\bn^{(1)}$ and  $\bn^{(2)}$.  One can take the normals such that
     the vectors $\bk$, $\bn^{(1)}$ and  $\bn^{(2)}$ will be noncoplanar. Then the vectors $\be_{1} =\bk\times\dfrac{\bk\times\bn^{(1)}}{|\bk\times\bn^{(1)}|}$ and $\be_{2} =\bk\times\dfrac{\bk\times\bn^{(2)}}{|\bk\times\bn^{(2)}|}$ in formula \eqref{qaxi2-1} for $\bK_-(\bn^{(1),(2)} ):\bq_+$ are different, and
     $\bK_-(\bn^{(1)} ):\bq_+\neq\bK_-(\bn^{(2)} ):\bq_+$. The rank reduction condition \eqref{case} is not fulfilled.

\item[(c)] The tensor $\bq_+$ is axisymmetric, and the eigenvalues $q$ and $q_k$ do not satisfy \eqref{qaxi1}. By \eqref{qaxi3}, \eqref{qaxi4}, only the case $$|q_k|<|q|$$ allows two different normals $\bn^{(1)}$ and  $\bn^{(2)}$, and both normals are perpendicular to the axe $\bk$.

 \item[(d)] The tensor $\bq_+$ is spherical, $\bq_+=q\bE$.  In the case of the rank-2 laminate this is just a particular case of the case $2(c)$ with $q_k=q$.

\end{itemize}

\emph{3. Rank-3 laminates}

Three  noncoplanar  vectors of normals have to satisfy the equalities
\begin{equation}\label{rank3}
\bq_+:\bK_-(\bn^{(1)}):\bq_+=\bq_+:\bK_-(\bn^{(2)}):\bq_+=\bq_+:\bK_-(\bn^{(3)}):\bq_+
\end{equation}
and provide the maximum of the quadratic form $\bq_+:\bK_-(\bn):\bq_+$. This cannot be the case if the eigenvalues of $\bq_+$ are different or if  the tensor $\bq_+$ is axisymmetric and eigenvalues $q$ and $q_k$ do not satisfy  \eqref{qaxi1} (by \eqref{qaxi3} and \eqref{qaxi4} only one normal $\bn=\bk$  or three coplanar normals lying in the  plane perpendicular to the axis of $\bq_+$ are possible).

Two cases remain:
\begin{itemize}
\item[(a)] The tensor $\bq_+$ is axisymmetric, the eigenvalues $q$ and $q_k$ satisfy \eqref{qaxi1}.  To satisfy the first equality in \eqref{rank3} we choose the normals $\bn^{(1)}$ and $\bn^{(2)}$  such that
    \begin{equation}\label{n-lam3}
    \bn^{(1)}=n_k^{(1)}\bk+n_1^{(1)}\be_1, \quad
    \bn^{(2)}=n_k^{(2)}\bk+n_2^{(2)}\be_2,
    \end{equation}
    where $\big(n_k^{(1)}\big)^{\!2}=\big(n_k^{(2)}\big)^{\!2}$ and
    $\big(n_1^{(1)}\big)^{\!2}=\big(n_2^{(2)}\big)^{\!2}$ are determined by \eqref{qaxi2},
    and $\be_1,\be_2,\bk$ is an arbitrary orthonormal basis related with the axis $\bk$. (To avoid misunderstanding note that subscripts 1, 2 (and $k$) denote here the component of a vector, not a number of a layer.)
        Then the second equality in \eqref{rank3} is fulfilled if
        \begin{gather}\label{Kq-lam3-1}
        \bn^{(3)}=n_k^{(3)}\bk+n_1^{(3)}\be_1,
        \end{gather}
        where
        $$n_k^{(3)}=n_k^{(1)}, \ n_1^{(3)}=-n_1^{(1)} \quad  \mbox{or}  \quad n_k^{(3)}=-n_k^{(1)}, \ n_1^{(3)}=n_1^{(1)},$$   {or} \begin{gather}
        \label{Kq-lam3-2}
        \bn^{(3)}=n_k^{(3)}\bk+n_2^{(3)}\be_2, \end{gather}
        where
        $$n_k^{(3)}=n_k^{(1)}, \ n_2^{(3)}=-n_2^{(2)} \quad \mbox{or} \quad
        n_k^{(3)}=-n_k^{(1)}, \ n_2^{(3)}=-n_2^{(2)}.$$

        By \eqref{qaxi2-1},
    \begin{gather*}
    \bK_-(\bn^{(1)}):\bq_+=\frac{q_k-q}{2\mu_-} \left(n_k^2\bk\otimes\bk-(1-n_k^2)\be_1 \otimes\be_1 \right),\\
    \bK_-(\bn^{(2)}):\bq_+=\frac{q_k-q}{2\mu_-} \left(n_k^2\bk\otimes\bk-(1-n_k^2)\be_2 \otimes\be_2 \right)
    \end{gather*}
    and
    \begin{equation}\label{Kq-lam3-3}
    \bK_-(\bn^{(1)}):\bq_+= \bK_-(\bn^{(3)}):\bq_+ \quad \mbox{or } \quad
     \bK_-(\bn^{(2)}):\bq_+= \bK_-(\bn^{(3)}):\bq_+
     \end{equation}
    in the cases \eqref{Kq-lam3-1} or  \eqref{Kq-lam3-2}, respectively.
    If to change $\bn^{(1)}$ or $\bn^{(2)}$ by  $\bn^{(3)}$ in relationships \eqref{n-lam3}--\eqref{Kq-lam3-3} then the equality
    \begin{equation}\label{Kq-lam3-4}
    \bK_-(\bn^{(1)}):\bq_+= \bK_-(\bn^{(2)}):\bq_+
     \end{equation}
    also becomes possible. Due to the equalities \eqref{Kq-lam3-3}, \eqref{Kq-lam3-4},  the rank can be lowered, the rank-3 laminate becomes equivalent to the rank-2 laminate if the tensor $\bq_+$ is axisymmetric. Thus, the case 3(a) is energy equivalent to the case 2(b).

\item[(b)] The tensor $\bq_+$ is spherical. Then any  three noncoplanar unit vectors can be the normals.
\end{itemize}
Therefore, there are four cases 1, 2(b),  2(c), 3(b) which represent all energy nonequivalent options to satisfy the conditions \eqref{min_n} and \eqref{min_beta} by the choice of the rank and directions of normals in dependence of the eigenvalues of the tensor $\bq_+$. In Section~3 this cases are I=1, II(a)=2(b), II(b)=2(c), and III =3(b).

\section{Representation of the translator in a form of the quadratic form \eqref{T}}
We note that the translator \eqref{translator} can written in the form of the convolution
\begin{equation}\label{varphit}
\varphi(\bt,\beps)=-\bt:\Cof\beps
\end{equation}
where the cofactor of the strain tensor is defined as
\begin{gather*}
\Cof\beps=\left\{
            \begin{array}{ll}
              (\det\beps)\beps^{-T}, & \hbox{if $\det\beps\neq 0$;} \\
              \eps_i\eps_j\be_k\otimes\be_k  \ (\mbox{no sum}), & \hbox{if $\eps_k=0,  \eps_i, \eps_j\neq 0 \ (i\neq j\neq k)$;} \\
              0, & \hbox{otherwise.}
            \end{array}
          \right.
\end{gather*}
where $\eps_i$ and $\be_i$ are eigenvalues and corresponding eigenvectors of $\beps$.

The convolution \eqref{varphit}, in turn, can be presented as a quadratic form \cite{AntimonovNTV-2010Bounds}
\begin{equation}\label{epsTeps}
	\bt:\Cof\beps=-\dfrac12\beps:\bT:\beps
\end{equation}
where the forth-rank translator tensor $\bT$ depends on $\bt$ as
\begin{equation}
\label{T1}
	\bT =(\bt\otimes\bE+\bE\otimes\bt)+(\tr\,\bt)(\bI-\bE\otimes\bE)-(\bI\cdot\bt+\bt\cdot\bI).
\end{equation}
The formulae \eqref{epsTeps}, \eqref{T1} can be easily derived if to take into account that if the inverse $\beps^{-1}$ exists then by the Cayley -- Hamilton theorem
 $$
\Cof\beps=\beps^2-I_1(\beps)\beps+I_2(\beps)\bE,
$$
where the strain invariants $I_1(\beps)=\tr\beps$ and $I_2(\beps)=\dfrac{1}{2}(I_1^2(\beps)-I_1(\beps^2))$ can be written as
$$
I_1(\beps)=\bE:\beps, \ I_2(\beps)=\frac{1}{2}(\beps:\bE\otimes\bE:\beps -\beps:\bI:\beps)
$$ and if to use the identities
\begin{gather*}
\bt:\beps^2\equiv\dfrac12\beps:(\bI\cdot\bt+\bt\cdot\bI):\beps,\\
 (\tr\beps)\beps:\bt=\beps:\bE\otimes\bt:\beps=\dfrac12 \beps:(\bE\otimes\bt+\bt\otimes\bE):\beps.
\end{gather*}

\section{Derivation for the phase transformation surfaces construction}

\emph{E1. Derivation of Eq. \eqref{Maxw1}}

Since $$
E=m_+(\gamma_*+\dfrac12 \be_0:\bM_-^{(n)}:\be_0),
$$
the derivative
\begin{equation}\nonumber
	\dfrac{\partial E}{\partial m_+} = \gamma_* + \dfrac12\be_0:\bM_-^{(n)}:\be_0+\dfrac12m_+\be_0:\dfrac{\partial \bM_-^{(n)}}{\partial m_+}:\be_0.
\end{equation}
By \eqref{Mn},
\begin{equation*}
\begin{split}
\dfrac{\partial \bM_-^{(n)}}{\partial m_+} &= -\bM_-^{(n)}: \dfrac{\partial}{\partial m_+}\left(\sk{\bC}^{-1}+(1-m_+) \sum\limits_{i=1}^{n}\beta^{(i)}\bK_-(\bn^{(i)})\right) : \bM_-^{(n)} \\
&=\bM_-^{(n)}: \left( \sum\limits_{i=1}^{n}\beta^{(i)}\bK_-(\bn^{(i)})\right) : \bM_-^{(n)}.
\end{split}
\end{equation*}

Then, taking into account the equality $\bM_-^{(n)}:\be_0 = \bq_+$  we get
\begin{equation}\label{derivation_1}
\begin{split}
	\dfrac{\partial E}{\partial m_+} &= \gamma_*+\dfrac12\bq_+:\left( \sk{\bC}^{-1}+(1-m_+) \sum\limits_{i=1}^{n}\beta^{(i)}\bK_-(\bn^{(i)}) + m_+\sum\limits_{i=1}^{n}\beta^{(i)}\bK_-(\bn^{(i)})\right):\bq_+\\
	&=\gamma_* +\dfrac12\bq_+:\left(  \sk{\bC}^{-1}+ \sum\limits_{i=1}^{n}\beta^{(i)}\bK_-(\bn^{(i)})   \right):\bq_+.
\end{split}
\end{equation}

By the optimality condition with respect to the  parameters $\beta^{(i)}$ and the normals $\bn^{(i)}$, $\bq_+:\bK_-(\bn^{(i)}):\bq_+=\mathcal{K}_{max}(\bq_+)$ at all normals $\bn^{(i)}$. Then \eqref{derivation_1} takes the form~\eqref{Maxw1}:
\begin{equation*}
	\dfrac{\partial E}{\partial m_+} = \gamma_*+\dfrac12(\bq_+: \sk{\bC}^{-1}:\bq_+ + \mathcal{K}_{max}(\bq_+)).
\end{equation*}

\emph{E2. Nucleation of the third-rank laminate}

In the case of the third-rank laminates $\bq_+=q\bE$. Eq. \eqref{Maxw1} takes form
\begin{equation}\nonumber
	2\gamma_*+q^2\left(\dfrac1{\sk k} + c\right)=0,
\end{equation}
where $c=3/(3k_-+4\mu_-)$.

Thus, the field in the phase ``$+$'' in third-rank laminates is a constant, defined by material parameters, and it is
\begin{equation}\label{3_rank_q}
	q=\pm\sqrt{-2\gamma_*\dfrac{\sk k}{1+c\sk k}}
\end{equation}

By \eqref{qplusM}, the relation between $\bq_+$ and $\bq_0$ is
\begin{equation}\nonumber
	\bq_0=q(\bE+c\sk{\bC}:\sum\limits_{i=1}^3\beta^{(i)}\bn^{(i)}\bn^{(i)})= q((1+c\sk{\lambda})\bE+2 c\sk\mu\sum\limits_{i=1}^3\beta^{(i)}\bn^{(i)}\bn^{(i)})
\end{equation}
at $m_-=1$.

The normals of the optimal third-rank laminate are directed along the principal directions of $\bq_0$. Thus,
\begin{equation}\label{strains_for_rank_3}
\begin{split}
	&q_{01}=q(1+c(\sk\lambda  +2\beta^{(1)}\sk\mu )), \\
	&q_{02} = q(1+c(\sk\lambda  +2\beta^{(2)}\sk\mu )),\\
	&q_{03} = q(1+c(\sk\lambda  +2\beta^{(3)}\sk\mu )).
\end{split}
\end{equation}

Summing these equations and using $\beta^{(1)}+\beta^{(2)}+\beta^{(3)}=1$ we get
\begin{equation}\nonumber
	\tr\bq_0=3q(1+c\sk k).
\end{equation}
and after substitution of  \eqref{3_rank_q} it becomes
\begin{equation}\label{3_rank_transform_surf}
	\tr\bq_0=\pm3\sqrt{-2\gamma_*\sk k \left(1+\sk kc\right)}.
\end{equation}

Eq. \eqref{3_rank_transform_surf} describes  a plane in the space of the eigenvalues of external strains. The transformation surface lays on the plane and it is restricted by the requirement $\beta^{(i)}\geq 0$. From \eqref{strains_for_rank_3} the structural parameters are
\begin{equation}\nonumber
	\beta^{(i)}= \dfrac1{2\sk\mu}\left( \dfrac1{c}\left(\dfrac{q_{0i}}{q}-1\right)-\sk\lambda  \right).
\end{equation}

With taking into account $\sk \mu>0$ and $c>0$ the restriction becomes
\begin{equation}\nonumber
	\dfrac{q_{0i}}{q}\geq 1+c\sk\lambda.
\end{equation}

\emph{E3. Nucleation of direct second-rank laminates}

By \eqref{qplusM}, the relation between $\bq_0$ and $\bq_+$ is
$$
\bq_0=\{\bI+m_- \sk\bC :(\beta \bK_-(\bn^{(1)})+(1-\beta)\bK_-(\bn^{(2)}))\}:\bq_+$$
where $\beta=m_1^{(1)}/m_-$. For the optimal laminates $\bk$ is one of the eigenvectors of $\bq_0$. Let $\bk=\be_1$. Then from \eqref{qaxi4} and \eqref{qaxi5} it follows at $m_-=1$ that
\begin{gather} \label{sk11}
	q_{01}=q_k+c\sk\lambda q, \\ \label{sk12}
	q_{02} = q+c(\sk\lambda q +2\beta\sk\mu \beta q),\\ \label{sk13}
	q_{03} = q+c(\sk\lambda q +2(1-\beta)\sk\mu q).
\end{gather}
From \eqref{sk12} \eqref{sk13} it follows that
\begin{equation}\label{sk14}
	q_{02}+q_{03} = 2q(1+c\sk{\lambda +\mu } ).
\end{equation}
Given $q_{01}$, $q_{02}$ and $q_{03}$, equations \eqref{sk11} and \eqref{sk14} determine $q$ and $q_k$ in dependence on $q_{01}$ and $q_{02}+q_{03}$. After substitution of these dependencies into \eqref{Maxw2} we obtain the equation of a cylindrical surface in space  $q_{01}$, $q_{02}$, $q_{03}$. The  nucleation surface is a part of the cylindrical surface restricted by the boundaries of $\Psi_2^q$ which correspond to $\beta=0$ and $1$ and $q_k=0$ and $q$.

\bibliographystyle{elsarticle-harv}
\bibliography{paper_bibliography}

\end{document}